\documentclass[%
 reprint,
superscriptaddress,
 amsmath,amssymb,
 aps,onecolumn
]{revtex4-2}

\usepackage[english]{babel}

\usepackage{amsmath}
\usepackage{graphicx}
\usepackage{setspace}
\usepackage{lineno}

\usepackage[colorlinks=true, citecolor=black, linkcolor=black, urlcolor=black]{hyperref}

\spacing{1.25}

\begin{document}

\title{Universal behavior of highly-confined heat flow in semiconductor nanosystems: from nanomeshes to metalattices}

\author{Brendan McBennett}
\affiliation{Department of Physics, JILA, and STROBE NSF Science and Technology Center, University of Colorado and NIST, Boulder, Colorado 80309}
\author{Albert Beardo}
\email{albert.beardo@colorado.edu}
\affiliation{Department of Physics, JILA, and STROBE NSF Science and Technology Center, University of Colorado and NIST, Boulder, Colorado 80309}
\author{Emma E. Nelson}
\affiliation{Department of Physics, JILA, and STROBE NSF Science and Technology Center, University of Colorado and NIST, Boulder, Colorado 80309}
\author{Bego\~{n}a Abad}
\affiliation{Department of Physics, JILA, and STROBE NSF Science and Technology Center, University of Colorado and NIST, Boulder, Colorado 80309}
\author{Travis D. Frazer}
\affiliation{Department of Physics, JILA, and STROBE NSF Science and Technology Center, University of Colorado and NIST, Boulder, Colorado 80309}
\author{Amitava Adak}
\affiliation{Department of Physics, JILA, and STROBE NSF Science and Technology Center, University of Colorado and NIST, Boulder, Colorado 80309}
\author{Yuka Esashi}
\affiliation{Department of Physics, JILA, and STROBE NSF Science and Technology Center, University of Colorado and NIST, Boulder, Colorado 80309}
\author{Baowen Li}
\affiliation{Department of Materials Science and Engineering, Department of Physics, Southern University of Science and Technology, Shenzhen 518055, PR China}
\affiliation{Department of Mechanical Engineering, Department of Physics, University of Colorado, Boulder, CO 80309, USA}
\author{Henry C. Kapteyn}
\affiliation{Department of Physics, JILA, and STROBE NSF Science and Technology Center, University of Colorado and NIST, Boulder, Colorado 80309}
\author{Margaret M. Murnane}
\affiliation{Department of Physics, JILA, and STROBE NSF Science and Technology Center, University of Colorado and NIST, Boulder, Colorado 80309}
\author{Joshua L. Knobloch}
\email{joshua.knobloch@colorado.edu}
\affiliation{Department of Physics, JILA, and STROBE NSF Science and Technology Center, University of Colorado and NIST, Boulder, Colorado 80309}

\begin{abstract}
Nanostructuring on length scales corresponding to phonon mean free paths provides control over heat flow in semiconductors and makes it possible to engineer their thermal properties. However, the influence of boundaries limits the validity of bulk models, while first principles calculations are too computationally expensive to model real devices. Here we use extreme ultraviolet beams to study phonon transport dynamics in a 3D nanostructured silicon \textit{metalattice} with deep nanoscale feature size, and observe dramatically reduced thermal conductivity relative to bulk. To explain this behavior, we develop a predictive theory wherein thermal conduction separates into a geometric \textit{permeability} component and an intrinsic \textit{viscous} contribution, arising from a new and universal effect of nanoscale confinement on phonon flow. Using experiments and atomistic simulations, we show that our theory applies to a general set of highly-confined silicon nanosystems, from metalattices, nanomeshes, porous nanowires to nanowire networks, of great interest for next-generation energy-efficient devices.

\end{abstract}

\maketitle

While electrons and spins are readily manipulated with external fields, no equivalent tool exists for phonons, which are the sole heat carriers in semiconductors. This lack of control over phonons has impeded efforts to engineer energy efficient devices with tailored thermal properties, including thermoelectrics and phononic analogs to electronic devices \cite{RevModPhys.84.1045}. One promising means of influencing phonons is through nanoscale structuring, where the introduction of features smaller than the mean free paths (MFPs) for phonon-phonon collisions impedes thermal transport, thereby modifying the thermal conductivity and providing greater control over heat flow \cite{NRM2021}.

Under highly confined heat transport conditions, phonons are usually assumed to be ballistic---implying that they travel between boundaries without internal collisions. Accordingly, the phonon MFPs are determined by the system size and geometry. Thermal conduction is commonly modeled in such situations using stochastic Monte Carlo solvers of the Boltzmann Transport equation, wherein phonons with velocities and intrinsic scattering rates sampled from bulk \textit{ab initio} calculations percolate through the nanoscale structure \cite{Hao2009}. This ballistic-stochastic approach has been applied extensively to explain dramatic reductions in thermal conductivity in 2D phononic crystals, consisting of periodic nanoscale voids in a crystalline background \cite{Hopkins2010, Heath2010}. However, geometric dependencies in these models must be computed case-by-case, and thus extrapolation to newly available 3D nanostructured systems is non-intuitive and computationally challenging \cite{Dabo2020,Nomura2022}.

Ballistic simulations predict a ray-like propagation of phonons in analogy with photons \cite{Anufriev2017}. This contrasts with recent Molecular Dynamics (MD) investigations of energy flow profiles in nanowires \cite{Verdier2019}, which show a Poiseuille-like flow reminiscent of hydrodynamics \cite{Melis2019,DESMARCHELIER2022123003}. These observations indicate the existence of a significant amount of phonon-phonon scattering and momentum diffusion at length scales much smaller than most of the bulk phonon MFPs, challenging the use of bulk phonon properties in nanoscale ballistic models \cite{JainPRB,Takahashi2020}. While hydrodynamic effects are known to be enhanced at low temperatures or in 2D materials, where most phonon-phonon collisions conserve momentum \cite{Cepellotti2015,ChenG2015}, increasing evidence suggests that they are also present at the nanoscale in semiconductors such as silicon--- even at room temperature \cite{Sendra2021, BeardoKnobloch2021,Guo2018,Zhou2017}. Therefore, a complete understanding of the breakdown of diffusive phonon transport in complex systems may require mesoscopic interpretations in analogy to fluid mechanics. Moreover, the degree to which heat dissipation through boundary scattering can be treated incoherently at the nanoscale remains unclear, since coherent effects modifying the phonon dispersion relations with respect to the bulk have been predicted and observed under some conditions \cite{LinaYang2017,Maire2017,Hussein2014,Hopkins2010}. These contradictions contribute to the lack of a unified description of heat transport in nanostructured semiconductors, especially at the smallest length scales and in emerging 3D phononic crystals \cite{Dabo2020}. 

In this work, we study highly-confined heat flow in a 3D silicon phononic crystal ``metalattice" at room temperature using time-resolved extreme ultraviolet (EUV) scatterometry. The metalattice periodicity is 36 nm, far below the average phonon mean free path in bulk silicon. We observe Fourier-like transport dynamics with an apparent thermal conductivity of only 1\% of bulk. To explain these and a broad set of similar measurements on highly-confined silicon nanosystems---including metalattices, nanomeshes, porous nanowires and nanowire networks \cite{Dabo2020,Heath2010,LinaYang2017,Takahashi2020}---we develop a predictive theory of heat flow in nanostructured silicon with dimensions far below the bulk phonon MFP. Through an analogy to the dynamics of rare gases in porous media, we decompose the thermal conductivity into permeability and viscosity components, where the former accounts for geometrical effects and the latter encompasses the intrinsic effect of nanoscale confinement on phonon flow and interactions. By comparing a broad set of experimental data with atomistic simulations, we uncover a universal relationship between viscosity and porosity, which yields a general analytical expression for thermal conductivity across highly-confined 2D and 3D silicon nanosystems. This surprising and general finding suggests a fundamentally modified phase space for heat transport enabled by phonon-phonon interactions in nanostructured silicon. Moreover, the resulting analytical model provides a route to predict phonon transport in boundary-dominated nanosystems for next-generation energy-efficient devices, making it possible for the first time to represent them as effective media in engineering applications.

\begin{figure}   \includegraphics[width=1.0 \columnwidth]{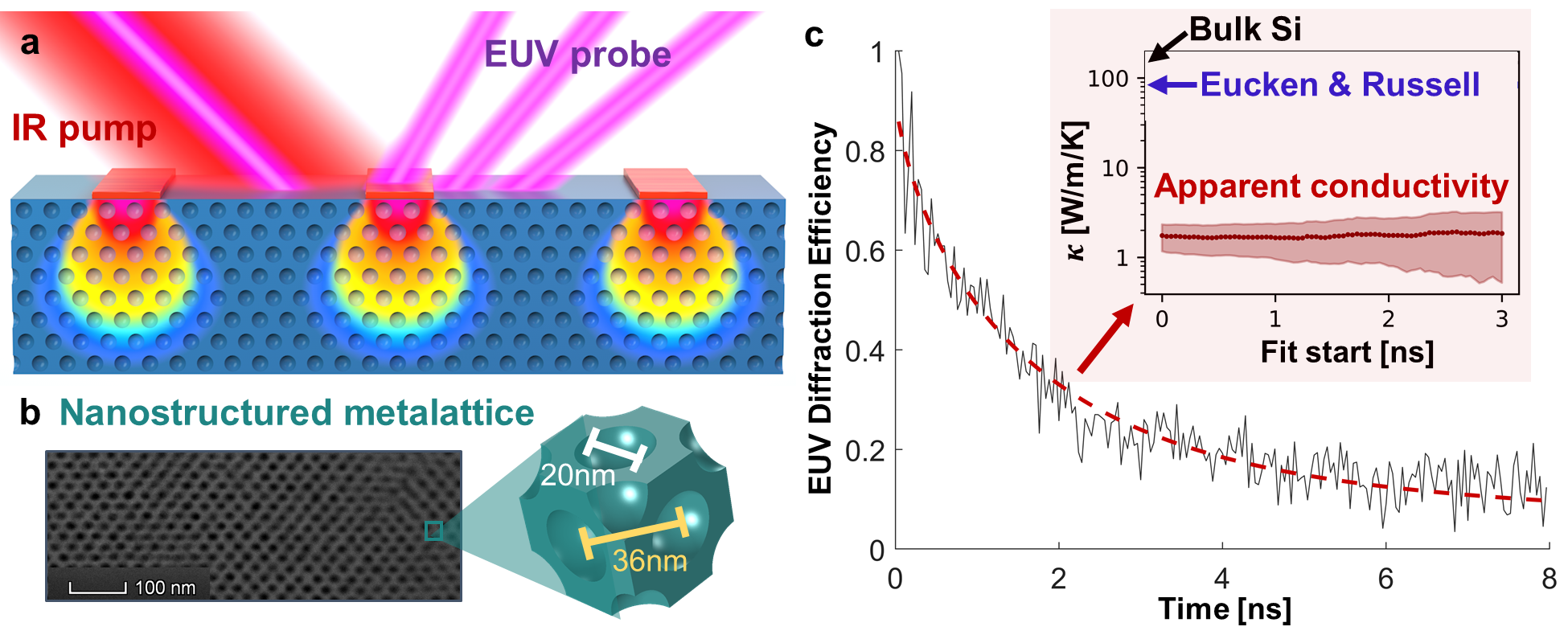}
    \caption{\textbf{Observation of ultra-low thermal conductivity in 3D silicon metalattices at room temperature.} (a) Schematic of the experimental setup. A time-delayed EUV probe monitors the thermal relaxation of a nickel grating on the metalattice surface after excitation by an ultrafast infrared laser pump pulse. (b) Cross-sectional electron microscopy image of the metalattice structure. The approximately 500 nm thick metalattice film consists of crystalline silicon punctured by FCC-packed pores of 36 nm periodicity and approximately 20 nm diameter, resulting in a porosity of 0.385 $\pm$ 0.02 \cite{Knobloch2022}. (c) An example experimental scan (gray) fit using Fourier's law (dashed red) with an ultra-low thermal conductivity. The inset plots the bulk silicon conductivity (black), volume-reduced Eucken and Russell thermal conductivity (blue) and the apparent metalattice thermal conductivity (red), which is only 1\% of the bulk. Note that since the apparent thermal conductivity does not vary with fit start time, the observed dynamics can be captured within a Fourier framework, where the metalattice is treated as an effective medium.}
    \label{fig:ML_data}
\end{figure}

The silicon metalattice film studied here consists of a crystalline silicon background interrupted by periodic FCC-packed spherical voids of approximately 20 nm diameter and 36 nm periodicity, as shown in Fig. \ref{fig:ML_data}(b). It is the identical sample investigated in \cite{Knobloch2022}, where its porosity and elastic properties were characterized in preparation for the present study. Details on the fabrication appear in \cite{Dabo2020}, \cite{Knobloch2022} and SI section 1. We excite the metalattice film using an ultrafast infrared pump laser and then monitor its time-delayed surface deformation using an ultrafast 30-nm-wavelength EUV probe \cite{Rundquist1998,Tobey2007}, as shown in Fig. \ref{fig:ML_data}(a). A set of 1D periodic nanoscale nickel gratings are fabricated on the sample surface, to act as both absorbers for the infrared pump and as a diffraction grating for the EUV probe. After excitation, the gratings relax by injecting energy across the nickel-silicon interface and into the metalattice film, as shown in Fig. \ref{fig:ML_data}(c). Further details on the experimental technique can be found in SI section 2. 

We model the thermoelastic response of the system using finite element methods and translate the simulated surface displacement to EUV diffraction efficiency, to directly compare with experimental data \cite{BeardoKnobloch2021} (details in SI section 3). Interestingly, the finite element simulations reveal that the transient relaxation of the metalattice can be accurately modeled, even at short timescales, using Fourier's law for the heat flux $\vec{q}$ and the temperature $T$,
\begin{subequations}
\begin{align}
\vec{q}=&-\kappa \nabla T,\label{FourierLaw}
\\
\nabla\cdot\vec{q}=&-(1-\phi)c_\text{V}\frac{\partial T}{\partial t},\label{energyconservation}
\end{align}
\end{subequations}
where $c_\text{V}$ is the bulk silicon specific heat, $\kappa$ is the apparent thermal conductivity of the effective medium representing the metalattice, and $\phi=0.385 \pm 0.02$ is the metalattice porosity from \cite{Knobloch2022}. The appropriateness of the Fourier model is demonstrated in the inset to Fig. \ref{fig:ML_data}(c), wherein the experimental dataset shows no systematic dependence of fitted conductivity on fit start time. A time-dependent thermal conductivity is a hallmark of non-Fourier phonon transport \cite{BeardoKnobloch2021}, because it indicates a thermal relaxation incompatible with Eq. \eqref{FourierLaw}, regardless of the assumed value of $\kappa$. The fits displayed in Fig. \ref{fig:ML_data}(c) constrain the apparent metalattice thermal conductivity to $\kappa=1.6 \pm 0.4$ W/mK (details in SI section 4), two orders of magnitude below the bulk crystalline value $\kappa_\text{bulk}=149$ W/mK. Contrary to previous studies on bulk silicon \cite{PNAS-hoogeboom,BeardoKnobloch2021}, we observe no meaningful variation in the fitted conductivity with heater linewidth and periodicity. The quality of the fits indicates that explicit simulation of the pores along with complex transport equations beyond effective Fourier's law is unnecessary. However, the measurements cannot be explained by the classical volume reduction effect, as described in the models of Eucken $\kappa_\text{Eucken}=\kappa_\text{bulk}(1-\phi)/(1+\phi/2)$ = 77 W/mK \cite{Song2004,Wang2006} or Russell $\kappa_\text{Russell}=\kappa_\text{bulk}(1-\phi^{2/3})/(1-\phi^{2/3}+\phi)$ = 82 W/mK \cite{Russell1935} and shown in Fig. \ref{fig:ML_data}(c). Therefore, a more fundamental explanation accounting for the interaction between the phonon population involved in the heat transport and the metalattice nanostructure is required.

To develop a comprehensive theory predicting the measured $\kappa$ in both the metalattice and a broader class of highly-confined nanostructured systems \cite{Dabo2020,LinaYang2017,Heath2010,Termentzidis2018}, we combine a phonon hydrodynamics framework \cite{Guo2015} and the theory of rarefied gas dynamics in porous media \cite{Carman1956,Alvarez2010}. Previous work has demonstrated that nanoscale size and frequency effects on heat transport in silicon can be modeled in terms of non-local and memory effects, as described by the Guyer-Krumhansl equation (GKE) \cite{Guyer1966}, with {\it ab initio} geometry-independent coefficients \cite{Sendra2021,BeardoKnobloch2021,RuizClavijo2021,Xiang2022}. However, the predictive capabilities of this hydrodynamic-like approach are restricted to systems with dimensions sufficiently large compared to the average bulk phonon MFP \cite{Beardo2019}, since nanoscale confinement modifies the {\it ab initio} phonon properties used to determine the GKE coefficients \cite{Melis2019, DESMARCHELIER2022123003}. This excludes the present metalattice and a broader class of nanostructured silicon materials. 

We overcome this limitation by using the similarities between the GKE and the Navier-Stokes equation for mass transport in fluids. In analogy to Stokes flow in incompressible fluids at low Reynolds numbers, we first simplify the GKE by assuming that, in highly-confined nanosystems, the thermal gradient is almost fully compensated by the viscous resistance described by the Laplacian term \cite{Alvarez2010}. Here, a highly-confined nanosystem refers to any crystalline semiconductor whose bulk average phonon MFP significantly exceeds the nanostructure length scale.

The resulting transport equation is a Stokes equation analog in the absence of volume viscosity effects:
\begin{equation}\label{Stokes}
\nabla T=\frac{\ell^2}{\kappa_\text{GK}}\nabla^2 \vec{q},
\end{equation}
where $\kappa_\text{GK}$ is the thermal conductivity appearing in the GKE and $\ell$ is an averaged resistive MFP known as the non-local length. In the previous equation, we identify $\mu\equiv\ell^2/\kappa_\text{GK}$ as the dynamic phonon gas viscosity, which models the strength of the interaction between the pore walls and the phonon population.

Inside the metalattice, Eq. \eqref{Stokes} can be further simplified in analogy to Darcy's law for fluid flow in porous media as
\begin{equation}\label{Darcy}
\nabla T=-\frac{\mu}{K}\ \vec{q},
\end{equation}
where $K$ is the metalattice permeability, which encompasses its geometrical properties. The permeability accounts for a non-zero heat flux, or slip heat flux, at the pore boundaries. This is analogous to the application of Darcy's law in rare gases, where collisions between fluid particles are scarce relative to momentum-destroying collisions with the boundaries \cite{Klinkenberg1941}. Details on the derivation of the Darcy's law analog from the GKE appear in the SI section 5.

The appropriateness of Eq. \eqref{Darcy} is validated by the success of the Fourier's law in Eq. \eqref{FourierLaw} in modeling the experiment, allowing the identification

\begin{equation}\label{interpretation_conductivity}
    \kappa=\frac{K}{\mu}.
\end{equation}
The apparent thermal conductivity can thus be interpreted as a ratio between a permeability $K$, which only depends on geometry, and the viscosity $\mu$, which is determined by the metalattice's averaged phonon dynamical properties.

This interpretation of the apparent conductivity enables the comparison of present and past experiments \cite{Dabo2020} and MD simulations \cite{Baowen2014} of metalattices with varying pore distributions, sizes, shapes and periodicities. Remarkably, it also enables comparison across a broad class of nanostructured silicon materials including 3D interconnected wire networks \cite{Termentzidis2018}, nanomeshes \cite{Heath2010}, and porous nanowires \cite{LinaYang2017}. The model is only valid for sufficiently constrained systems, with characteristic sizes of a few tens of nanometers. It cannot be extended to larger-scale  structured materials (\textit{e.g.}, \cite{PeidongYang2017,Hopkins2010}) at room temperature due to the breakdown of the Stokes regime assumption neglecting the heat flux contribution in the GKE (details in SI subsection 6B).

In addition to quantifying $\kappa$ in the different systems, it is necessary to characterize the permeability $K$---a purely geometrical quantity---to extract the phonon viscosity $\mu$. We follow the Kozeny-Carman derivation \cite{Carman1956}, including the Klinkenberg quasiballistic correction and assuming diffusive phonon-boundary collisions \cite{Klinkenberg1941} (see SI section 5), to obtain the following expression:
\begin{equation}\label{permeability}
    K=\frac{d^2\Psi^2(1-\phi)^3}{24\Gamma^2\phi^2},
\end{equation}
where $d$ is the average empty pore diameter, $\Psi$ is the sphericity of the pores ({\it i.e.}, the surface area ratio of the ideal spherical pore to the actual pore), and $\Gamma$ is the tortuosity ({\it i.e.}, the ratio of the average length of a heat flux streamline to the distance between its ends). Then, using Eqs. \eqref{interpretation_conductivity} and \eqref{permeability}, one can characterize the phonon viscosity $\mu$ in systems with a 3D distribution of empty pores using only the measured apparent conductivity $\kappa$ and the average geometric properties ($\phi,d,\Psi,\Gamma$). Expression \eqref{permeability} can be easily adapted to characterize permeability in a variety of geometries beyond metalattices. For example, a nanowire network can be modeled as a porous medium, with porosity and pore size defined in terms of wire diameter and periodicity, while for nanomeshes, the permeability can be expressed in terms of a 2D distribution of cylindrical pores in a thin film (see SI subsection 5B and Eq. S24). The sphericity parameter $\Psi$ can be characterized via electron microscopy to model imperfectly spherical pores in experiments, or used to account for other pore shapes such as cubes.

\begin{figure*}
    \includegraphics[width=1.0 \columnwidth]{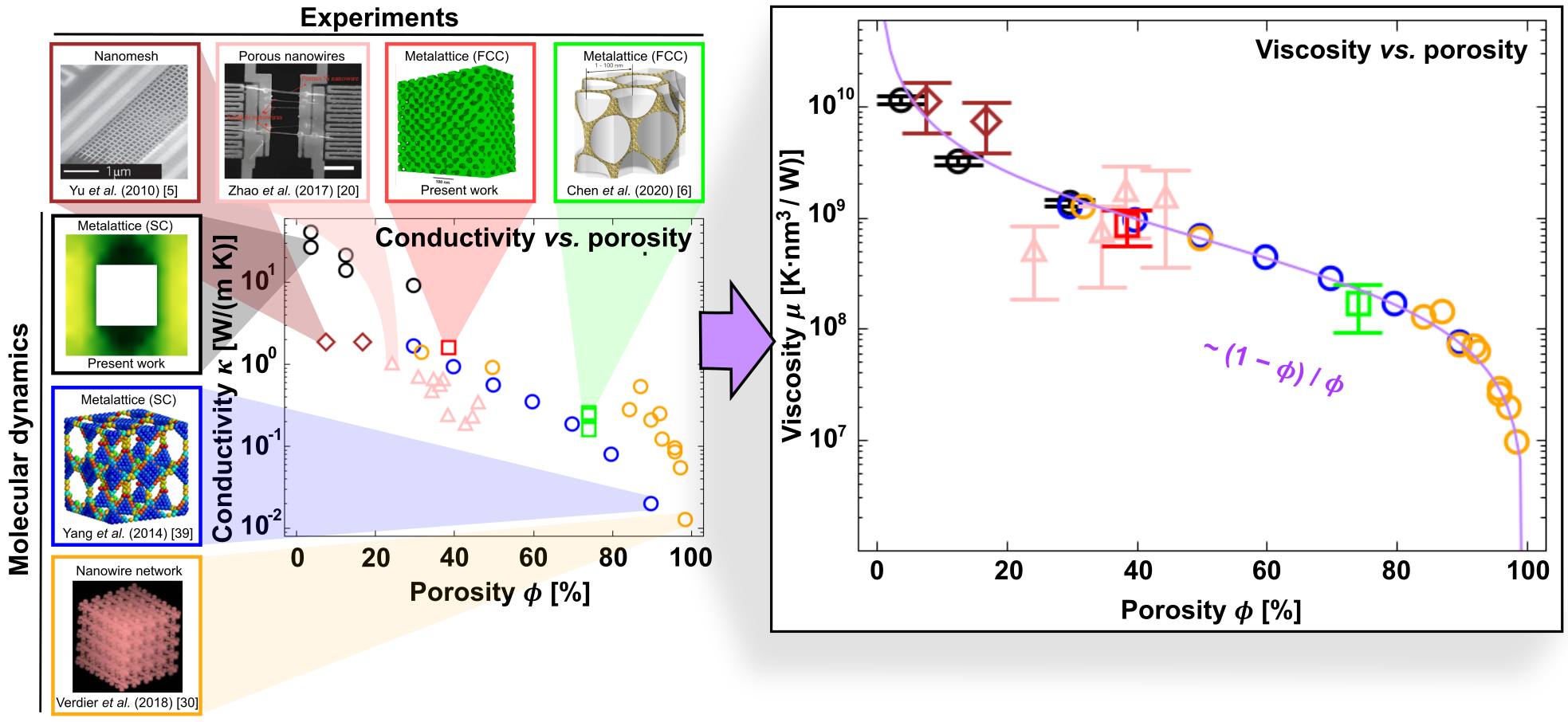}
\caption{\textbf{Universal scaling of viscosity with porosity in highly-confined nanosystems}. Left: Apparent thermal conductivity $\kappa$. Right: Viscosity $\mu$ from Eq. \eqref{interpretation_conductivity} as a function of nanostructured silicon system porosity. The viscosity captures the intrinsic phonon properties in nanostructured silicon systems regardless of their specific geometry, and follows a universal trend. The theoretical results considered include new MD simulations of square-pore metalattices (black circles), along with previous MD simulations of circular-pore metalattices (blue circles---inset reproduced from \cite{Baowen2014}. Copyright 2014 ACS) and 3D nanowire networks (yellow circles---inset reprinted with permission from \cite{Termentzidis2018}. Copyright 2018 APS). Experimental results considered include data on circular-pore FCC-packed metalattices from the present study (red square), a high-porosity metalattice with similar geometry from a previous study (green square---inset reproduced from \cite{Dabo2020}. Copyright 2014 ACS), along with previous data on 2D cylindrical-pore nanomeshes (brown diamonds---inset reproduced from \cite{Heath2010}. Copyright 2010 Springer Nature) and porous nanowires (pink triangles---inset reproduced from \cite{LinaYang2017}. Copyright 2017 WILEY-VCH Verlag GmbH). In all experiments, phonon transport is confined in 3D to dimensions approximately 10\% or less of the average phonon mean free path in bulk silicon. All details on the sample geometry and permeability calculations in each case appear in SI subsection 5C.}
    \label{fig:kappa_KVSporosity}
\end{figure*}

In Fig. \ref{fig:kappa_KVSporosity}, we show the apparent conductivity $\kappa$ and the viscosity $\mu$ as functions of porosity for new and previous MD simulations in metalattices \cite{Baowen2014} and 3D nanowire networks \cite{Termentzidis2018}, along with present and past experimental results in metalattices \cite{Dabo2020}, nanomeshes \cite{Heath2010}, and porous nanowires \cite{LinaYang2017} (details in SI sections 5C and 6). The thermal conductivity $\kappa$ at a given porosity depends strongly on specific geometrical details such as pore size, shape and distribution. Conversely, the viscosity $\mu$ depends solely on porosity across all nanostructured systems in both experiments and MD simulations. More precisely, the viscosity scales as $(1-\phi)/\phi$, the ratio of the available to non-available volume for heat transport. Combining this universal scaling with Eqs. \eqref{interpretation_conductivity} and \eqref{permeability}, one obtains an analytical expression for the apparent thermal conductivity in nanostructured silicon systems with feature sizes far below the average bulk phonon MFP. 

Since the intrinsic phonon properties are seen to depend primarily on nanosystem porosity, regardless of the specific geometry, the variations in thermal conductivity at a given porosity observed in Fig. \ref{fig:kappa_KVSporosity} can be attributed to the permeability $K$, which only depends on geometry. Permeability not only accounts for the classical volume reduction effect, but also captures nanoscale geometrical effects, and can be used to reinterpret correlations between thermal conductivities and geometric descriptors observed in the literature \cite{Zianni2014,WEI2020120176,D2CP00775D,Lim2016}. For example, the differences between the permeability and the surface-to-volume ratio justify why the latter is not a general proxy for thermal conductivity in nanostructured systems \cite{Heath2010}. The influence of the neck size on the permeability can be investigated by identifying the neck size with the channel diameter $D$ and using the 3D geometrical relation, $D = 2(1-\phi)\Psi d/3\phi$ in Eq. \eqref{permeability}, as discussed in SI section 5A. The corresponding 2D geometrical relation can be found in SI section 5B, Eq. S23. The neck size has been shown to be a good indicator of the effects of nanostructuring on thermal conductivity \cite{Nomura2022,Anufriev2016}, and the present model agrees with previous findings that thermal conductivity varies as the square of neck size under highly-confined transport conditions \cite{Lim2016}. Moreover, some purely geometrical effects commonly invoked in phononic crystals like pore anisotropy \cite{Ferrando2018}, backscattering \cite{PeidongYang2017}, or surface disorder \cite{Pierre2009} can be related to the permeability. Pore anisotropy might be described by the tortuosity $\Gamma$, and surface disorder may modify  the slip flux boundary condition, which determines the numerical factor in Eq. \eqref{permeability} (details in SI section 5). However, it is worth noting that it was not necessary to characterize the influence of these effects when calculating each sample's permeability in order to uncover the universal scaling of $\mu$, suggesting that these are not the key mechanisms distinguishing the different cases considered. Finally, the resistive effect associated with backscattering in ballistic interpretations is quantified by the geometrical parameters appearing in $K$, which depend on the size and frequency of channels orthogonal to the thermal gradient.

The universal relationship between viscosity and porosity is reminiscent of a non-Newtonian fluid \cite{Chhabra2010}, and provides insight into the intrinsic mechanisms influencing phonon transport in highly-confined nanosystems. According to kinetic theory \cite{Sendra2021}, the thermal conductivity $\kappa_\text{GK}$ and non-local length $\ell$ are related to the phonon MFP $\Lambda$ as $\kappa_\text{GK}\propto \Lambda$ and $\ell^2\propto \Lambda^2$ respectively. This implies that $\mu\propto \Lambda$. However, in Fig. \ref{fig:kappa_KVSporosity}, systems with the same porosities but different neck sizes are shown to have the same viscosity. In line with previous work, this contradicts the ballistic assumption wherein $\Lambda$ is determined by the neck size, and challenges the applicability of the standard Matthiessen's rule in estimating scattering times \cite{Cartoixa2016,HOSSEINI2022100719}. In the fully ballistic limit where porosity approaches 100\%, $\mu$ declines to zero. However, we observe that even at high porosities above 80\% ($\phi$ = 0.8), MD simulation geometries with dramatically different values of $\kappa$ collapse to the same value of $\mu$, indicating that a hydrodynamic-like description of the phonon flow can be extended to extremely constrained systems. These results suggest that the phonon distribution evolves collectively under nanoscale confinement, and should be characterized by averaged phonon properties including $\ell$, $\kappa_\text{GK}$ and $\mu$, rather than multiscale descriptions where each phonon mode evolves independently with a distinct MFP. This perspective is consistent with recent electron transport experiments, where current profiles exhibit hydrodynamic-like vortices at low temperatures \cite{Steinberg22}. Since momentum-conserving electron-electron collisions are scarce under these conditions, the observed behavior is instead induced by nanoscale confinement, in analogy to the emergence of phonon hydrodynamics in silicon nanosystems. 

\begin{figure}
    \centering
    \includegraphics[width=0.7 \columnwidth]{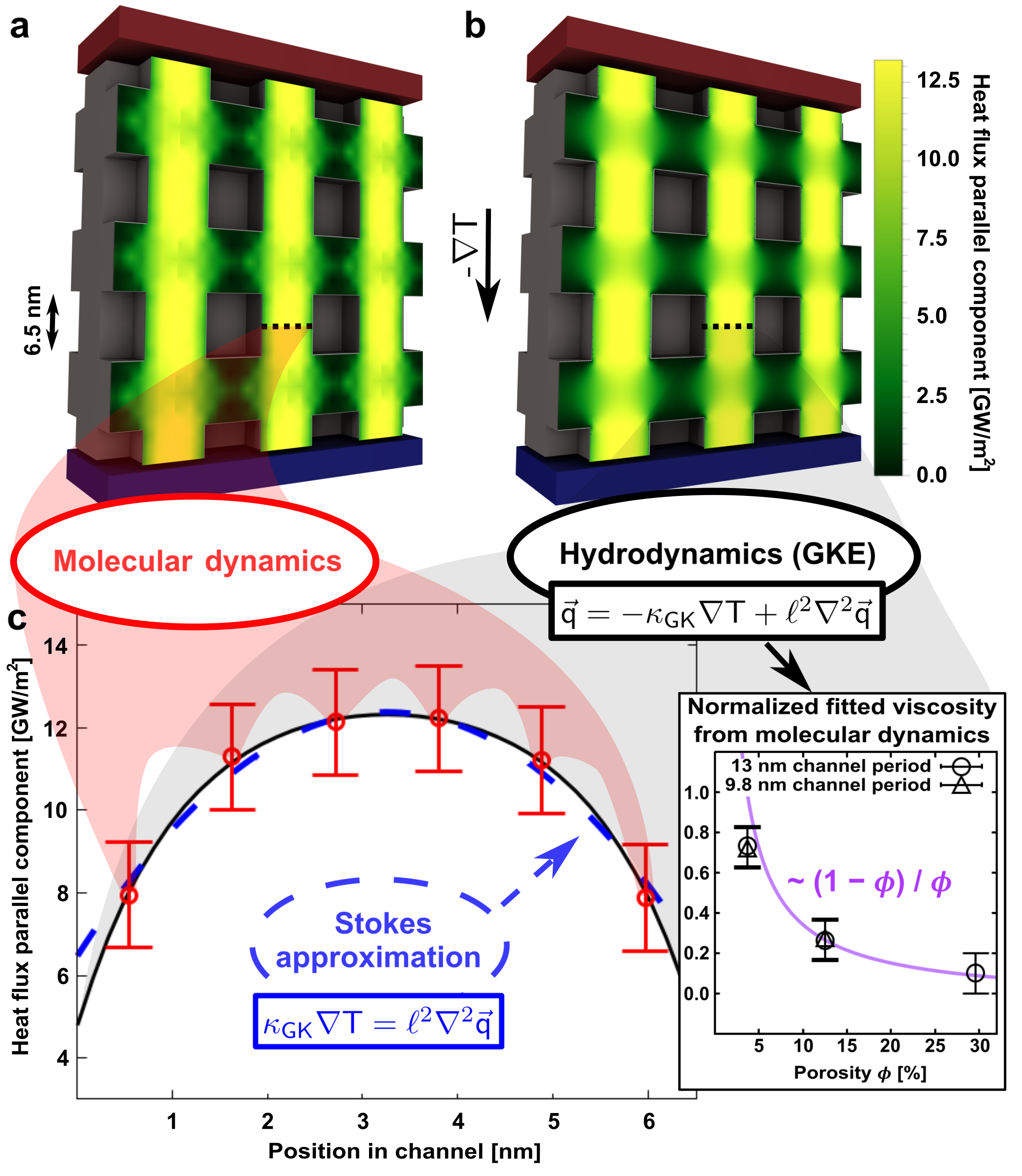}
    \caption{\textbf{Fluid-like heat flow in highly-confined nanosystems.} Parabolic heat flux profiles in 3D metalattices are predicted by both molecular dynamics simulations and hydrodynamic simulations, confirming the universal viscosity scaling. (a) Heat flux profile according to an MD simulation of a metalattice with cubic pores (see SI section 6). The channel spacing is 13 nm and the pore side length is 6.5 nm. The imposed temperature difference between the ends of the system is 20 K. (b) Heat flux profile fitted to the same geometry using the GKE and appropriate parameter values for $\ell$ and $\kappa_\text{GK}$. (c) Heat flux profile along a cross-section between two contiguous pores according to MD (dots) and the GKE fit (solid black line). The Stokes fit (dashed blue line) can locally reproduce the flux in the passage between the pores. The inset plots the viscosity $\mu=\ell^2/\kappa_\text{GK}$ obtained from the GKE fits for channel periods of 13 nm (circles) or 9.8 nm (triangles). It follows the same scaling with porosity as observed in experiments.}
    \label{fig:Fig3}
\end{figure}

To investigate the universal scaling of $\mu$, we directly quantify its magnitude using local heat flux profiles calculated from atomistic simulations. In Fig. \ref{fig:Fig3}, we show the steady-state heat flux profile in a 3D metalattice as obtained from MD simulations (details in SI section 6). We then fit the atomistic profiles using the steady-state GKE 
\begin{equation}\label{GKE}
    \vec{q}=-\kappa_\text{GK}\nabla T+\ell^2\nabla^2\vec{q}
\end{equation}
with appropriate values for $\kappa_\text{GK}$ and $\ell$, and slip boundary conditions. The excellent fit quality further validates the use of hydrodynamic-like modeling to describe heat transport in highly-confined nanosystems. The Poiseuille-like profiles obtained in MD are incompatible not only with the exponential profiles predicted by the Fuchs-Sondheimer approximation to the Boltzmann transport equation \cite{DESMARCHELIER2022123003}, but also with more general ballistic descriptions using the bulk MFP spectrum \cite{Anufriev2017}. The viscosity fit with the steady-state GKE $\mu=\ell^2/\kappa_\text{GK}$ follows the same scaling with porosity obtained from experiments using $\kappa$ and $K$ in Eq. \eqref{permeability}. The observed decrease in $\mu$ with increasing porosity is due to a reduction in the size of strongly correlated heat flux regions surrounding the boundaries, whose extent is described by $\ell$. However, the conductivity parameter $\kappa_\text{GK}$ is also slightly reduced. For fixed porosity, both $\ell$ and $\kappa_\text{GK}$ increase when increasing the pore spacing, but the resulting viscosity $\mu$ remains the same. In Fig. \ref{fig:Fig3}, we also show that the Stokes approximation in Eq. \eqref{Stokes} can fit the local heat flux profile in the passages between contiguous pores, as required in the Darcy's law derivation. However, the Stokes approximation is not sufficient to predict the complete heat flux profile in MD simulations, including the complex pore boundaries (details on the GKE and Stokes fits in SI subsection 6B).

Since the volume reduction effect and the geometrical influence of the boundaries on thermal conductivity are included in the permeability, the viscosity $\mu$ might be related to intrinsic phonon properties and the emergence of coherent or incoherent phonon effects in nanosystems. For example, previous work experimentally observed elastic softening leading to reduced phonon velocities with increasing porosity, regardless of the specific geometric details \cite{LinaYang2017, Takahashi2020}. Elastic softening is not governed by the permeability and may influence the relationship between viscosity and porosity uncovered in Fig. \ref{fig:kappa_KVSporosity}. Another possible source of variation in $\mu$ is the hypothetical presence of amorphous layers surrounding the pores in the sample, which are not considered in the Fig. \ref{fig:kappa_KVSporosity} MD data. Their presence might also reduce heat flux in the crystalline channels, which would result in an increased viscosity \cite{DESMARCHELIER2022123003,Lysenko1999}. Moreover, we find that viscosity is independent of disorder in the nanostructured system, {\it e.g.}, the periodically arranged 3D metalattices display the same viscosity as the randomly distributed pores in the nanowires. Hence, coherent interference effects previously reported in periodically structured silicon systems at very low temperatures \cite{Zen2014,Maire2017} are negligible in the mainly room temperature experiments and simulations considered here. Finally, previous work detected the localization of phonon modes within metalattices at very high porosities \cite{Baowen2014}, which directly influences the total contribution of the phonon population to $\ell^2$, $\kappa_\text{GK}$, and $\mu$. A deeper understanding of the universal behavior uncovered here might require relating these average phonon properties with the dispersion relations in nanostructured silicon systems or with the modification of the phonon-phonon scattering rates in nanostructured systems relative to bulk \cite{PNAShossein}.

In conclusion, we show that it is possible to use an effective medium theory with a Fourier relation to model heat transport in 3D metalattices and other highly-confined silicon nanosystems. In contrast to a ballistic interpretation where phonon modes evolve independently, the present approach stems from an analogy to rarefied fluid flow in porous media, motivated by Poiseuille-like heat flux profiles observed in atomistic simulations of nanoscale channels. In highly-confined nanosystems, thermal conduction separates into a permeability component, which captures the geometrical effect of complex structures, and a viscosity component related to the intrinsic phonon properties in the nanosystem environment. Variations in thermal conductivity between experiments or MD simulations at constant porosity can be attributed to the permeability, whereas viscosity depends solely on porosity and varies universally as $(1-\phi)/\phi$. This leads to an analytical description of thermal conduction and suggests that the viscosity is the appropriate metric for exploring the modification of the phonon dispersion relations and scattering rates in semiconductor nanosystems beyond the ballistic interpretation. In general, these ideas might apply to describe the transport of other particles beyond phonons, where hydrodynamic signatures emerge in the presence of slowly evolving magnitudes that violate the local equilibrium assumption \cite{Varnavides2020,Cheung2019,Gromov2020}. 

\section*{Acknowledgements}

The authors gratefully acknowledge support from the STROBE National Science Foundation Science \& Technology Center, Grant No. DMR-1548924. A.B. acknowledges support from the Spanish Ministry of Universities through a Margarita Salas fellowship funded by the European Union - NextGenerationEU. This work utilized the Alpine high performance computing resource at the University of Colorado Boulder. Alpine is jointly funded by the University of Colorado Boulder, the University of Colorado Anschutz, and Colorado State University. The authors would also like to acknowledge Dr. Guimei Zhu for her proofing of the manuscript and Steven Burrows for assistance designing Fig. \ref{fig:ML_data}(a). Finally, we note Amitava Adak's current affiliation in the Department of Physics, IIT (ISM) Dhanbad, Jharkhand 826004, India and Bego\~{n}a Abad's current affiliation in the Physics Department, University of Basel, Klingelbergstrasse 82, CH-4056 Basel, Switzerland.

\section*{References}

\bibliographystyle{ieeetr}
\bibliography{biblio}

\end{document}


\title{SUPPLEMENTARY INFORMATION\\
Universal behavior of highly confined heat flow in semiconductor nanosystems: from nanomeshes to metalattices}

\author{Brendan McBennett}
\affiliation{Department of Physics, JILA, and STROBE NSF Science and Technology Center, University of Colorado and NIST, Boulder, Colorado 80309}
\author{Albert Beardo}
\email{albert.beardo@colorado.edu}
\affiliation{Department of Physics, JILA, and STROBE NSF Science and Technology Center, University of Colorado and NIST, Boulder, Colorado 80309}
\author{Emma E. Nelson}
\affiliation{Department of Physics, JILA, and STROBE NSF Science and Technology Center, University of Colorado and NIST, Boulder, Colorado 80309}
\author{Bego\~{n}a Abad}
\affiliation{Department of Physics, JILA, and STROBE NSF Science and Technology Center, University of Colorado and NIST, Boulder, Colorado 80309}
\author{Travis D. Frazer}
\affiliation{Department of Physics, JILA, and STROBE NSF Science and Technology Center, University of Colorado and NIST, Boulder, Colorado 80309}
\author{Amitava Adak}
\affiliation{Department of Physics, JILA, and STROBE NSF Science and Technology Center, University of Colorado and NIST, Boulder, Colorado 80309}
\author{Yuka Esashi}
\affiliation{Department of Physics, JILA, and STROBE NSF Science and Technology Center, University of Colorado and NIST, Boulder, Colorado 80309}
\author{Baowen Li}
\affiliation{Department of Materials Science and Engineering, Department of Physics, Southern University of Science and Technology, Shenzhen 518055, PR China}
\affiliation{Department of Mechanical Engineering, Department of Physics, University of Colorado, Boulder, CO 80309, USA}
\author{Henry C. Kapteyn}
\affiliation{Department of Physics, JILA, and STROBE NSF Science and Technology Center, University of Colorado and NIST, Boulder, Colorado 80309}
\author{Margaret M. Murnane}
\affiliation{Department of Physics, JILA, and STROBE NSF Science and Technology Center, University of Colorado and NIST, Boulder, Colorado 80309}
\author{Joshua L. Knobloch}
\email{joshua.knobloch@colorado.edu}
\affiliation{Department of Physics, JILA, and STROBE NSF Science and Technology Center, University of Colorado and NIST, Boulder, Colorado 80309}

\maketitle
\tableofcontents

\section{Metalattice Sample Fabrication and Sphericity Characterization}\label{Fabrication}

The silicon metalattice sample studied in this work is the same sample studied in \cite{Knobloch2022}, where its porosity, elastic properties and thickness were characterized with a combination of the extreme ultraviolet (EUV) scatterometry technique described in Section \ref{sectionEUV} and electron tomography. Use of the identical sample provides confidence that the material parameters in the finite element analysis (FEA) model used to fit the apparent thermal conductivity are correct. The fabrication of this metalattice sample from silica nanosphere template synthesis \cite{Hartlen2008, Watanbe20111} and deposition \cite{Kuai2004, Russell2017}, through silicon infiltration, annealing and silica etching \cite{Liu2018, Dabo2020, Abad2020, Knobloch2022} is documented in the literature.

The sphericity is a key quantity in the computation of the permeability and is a required parameter for good agreement between the experiment and the empirical viscosity behavior shown in Fig. 2 of the main text. The sphericity, $\Psi$, as defined in Section \ref{Permeability}, is the ratio, or correcting factor, between surface area of a perfectly spherical pore of equivalent volume and the experimental pore surface area. If the pores in the measured metalattice were precisely spherical, then $\Psi = 1$; however, as shown in our previous work \cite{Knobloch2022}, this is not the case, thus $\Psi < 1$. Moreover, the pores in the experimental metalattice contain interconnections which are not directly incorporated into the theoretical permeability calculation for the metalattice. However, by measuring the sphericity, we can precisely account for the irregular pore shape and their interconnections. The segmented electron microscopy tomography data from our previous work \cite{Knobloch2022} provides the nanoscale 3D structure of the measured metalattice sample, allowing us to extract the sphericity. To do this, we compute an idealized theoretical surface area to volume ratio, $S_\text{ideal} / V_\text{ideal}$, for a unit cell of the metalattice geometry and divide it by the equivalent experimental ratio extracted from the tomography data set, $S_\text{tomo} / V_\text{tomo}$. Therefore, the sphericity will be given by:
\begin{equation}
\label{expsphericity}
    \Psi = \frac{S_\text{ideal}}{V_\text{ideal}} \frac{V_\text{tomo}}{S_\text{tomo}},
\end{equation}

As shown in our previous work, the porosity of the metalattice is $\phi=0.385 \pm 0.02$, while the side length of the metalattice unit cell is $a = 36$ nm \cite{Knobloch2022}. The theoretical pore diameter, $d$, can be simply related to the porosity and the unit cell size via: $\phi=\left(^{16\pi}/_{3a^3}\right)\left(^d/_{2}\right)^3$. Therefore, using the FCC pore distribution, $S_\text{ideal}$ is simply 4 times the surface area of a sphere with diameter $d$. Since $V_{\text{tomo}}$ is easily calculated from the size of the tomography data set, the remaining parameter to calculate is $S_\text{tomo}$. We utilize image processing tools implemented in {\it MATLAB} \cite{matlab} to compute the total surface area of the segmented electron microscopy tomography data set (see Fig. 4 in Ref. \cite{Knobloch2022}) and subtract the surface area of the external boundaries from that value which results in a value for the internal surface area only. We note that errors could arise due to digitization of the internal surfaces of the metalattice since the tomography resolution is 1.4 nm which is on the same scales as the pore diameter $\sim$20 nm. We therefore qualify that our image processing algorithm always underestimates the surface area, thus our sphericity calculation is an upper bound. However, we expect this error to be $<5\%$ since larger errors are only incurred on a few voxel pore radii. Using these calculations and Eq. \eqref{expsphericity}, we find that $\Psi$=0.3614.  

We perform an analogous analysis on the nanomesh experiments from Ref. \cite{Heath2010}. However, rather than computing surface areas of spheres, we calculate the ratio between the cross-sectional areas of experimental and idealized cylindrical pores (that is, compute the circularity of the pores from the 2D images). Often the provided electron microscope images of the measured samples do not contain high enough resolution to accurately compute the sphericity. Therefore, we interpolate the images before computing the sphericity. To ensure our interpolation method does not affect the resulting number, we use two different methods and average the results. One method preserves the pixelated shape (interpolated points are assigned values equal to their nearest neighbor) while another artificially smooths the image (cubic spline interpolation). Moreover, we ensure that the sphericity is converged with respect to the number of interpolation points. We calculated that $\Psi \sim 0.75$ for the nanomeshes.

\section{Extreme Ultraviolet Scatterometry Measurements}
\label{sectionEUV}
We use an EUV scatterometry experiment to obtain the data used to fit the silicon metalattice thermal conductivity. In order to excite thermal and acoustic responses in the metalattice, first 1D nickel gratings of varying linewidth and periodicities are fabricated on top of the metalattice film using an electron beam lithography process described in Ref. \cite{Abad2020}. These nickel gratings are then excited using $\sim$50 fs pulses from a 4 kHz Ti:Sapphire amplifier, causing them to heat and expand, launching coherent acoustic waves and injecting heat into the metalattice below \cite{Abad2020, Frazer2019}. Unlike visible laser experiments, EUV light is insensitive to changes in the refractive index induced by valence electrons, allowing us to directly track the sample's surface displacement with a time-delayed EUV probe as it cools \cite{Tobey2007}. The EUV probe is created via high harmonic generation, an extreme nonlinear process that upconverts photons from the infrared Ti:Sapphire amplifier into higher harmonics of much shorter wavelength. The experimental setup in this work uses an argon-filled waveguide to generate the 25th through 31st harmonics, resulting in EUV pulses with a central wavelength of approximately 29 nm \cite{Rundquist1998}. 

The EUV probe beam is diffracted from the nickel gratings and detected on an EUV-sensitive CCD camera. By comparing the EUV diffraction patterns with and without the infrared pump excitation, it is possible to detect modulations caused by the heating of the nickel gratings and the surface acoustic waves in the material. The 29 nm central wavelength and 10 fs pulse duration of the EUV probe provide excellent spatio-temporal resolution, including sub-angstrom sensitivity to grating height changes, and allow the investigation of nickel grating geometries with periodicities far below the visible diffraction limit \cite{Hernandez-Charpak2017}. The technique is also noncontact and nondestructive, making it useful for many types of experimental applications. 

\section{Metalattice Finite Element Model}
\label{MLFEA}

\begin{figure}
    \centering
    \includegraphics[width=0.9 \columnwidth]{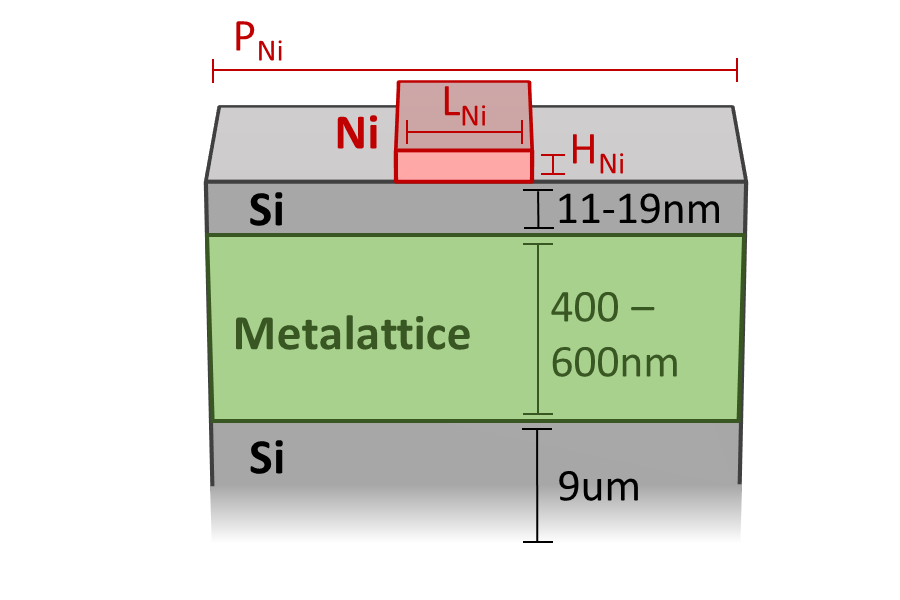}
    \caption{\textbf{Finite element model geometry.} Simulated geometry of the nickel nanoheaters, metalattice film and silicon substrate. The silicon layer directly underneath the nickel is confirmed by cross-sectional SEM and results from the silicon re-infiltration step in the metalattice fabrication.}
    \label{fig:FigureS1}
\end{figure}

Implemented in the COMSOL Multiphysics software \cite{comsol}, the metalattice FEA model represents one period of the nickel grating structure, metalattice film, and silicon substrate, as shown in Fig. \ref{fig:FigureS1}. Insulating boundary conditions apply on the exposed surface, since radiation losses in vacuum are negligible on the timescales considered. Periodic boundary conditions apply on the left and right edges. Since modeling the entire 100s of $\upmu$m substrate is computationally impractical, we include only the first 9 $\upmu$m and verify that this depth is sufficiently large that the simulation results are unaffected over the timescales considered. We model the metalattice itself as a homogeneous silicon layer with modified material properties rather than a network of pores. Having previously determined its average porosity and elastic properties \cite{Knobloch2022}, the metalattice density, specific heat capacity, Young's modulus and Poisson's ratio are taken as inputs; only the apparent conductivity is fit. The metalattice film, nickel grating and silicon substrate material properties are all summarized in Table \ref{tab1}.

\begin{table}[hbt!]
\centering
\begin{tabular}{ |p{1.8cm}|p{3.4cm}|p{1.3cm}| }
 \hline
 \textbf{Parameter} & \textbf{Value} & \textbf{Source} \\
 \hline
 $\rho_\text{Ni}$ & 8.9 g/cc & \cite{Lide2005} \\
 $c_\text{p Ni}$ & 444 J/kg/K & \cite{Lide2005} \\
 $\kappa_\text{Ni}$ & 90.9 W/m/K & \cite{Lide2005} \\
 $\alpha_\text{Ni}$ & 12.77e-6 $K^{-1}$ & \cite{Kollie1977} \\
 $E_\text{Ni}$ & 219 GPa & \cite{Nardi2011} \\
 $\nu_\text{Ni}$ & 0.31 & \cite{Nardi2011} \\
 \hline
 $\phi$ & 0.385 $\pm$ 0.02 & \cite{Knobloch2022} \\
 $\rho_\text{ML}$ & 2.33(1-$\phi$) = 1.43 g/cc & \cite{Lide2005} \\
 $c_\text{p ML}$ & 712 J/kg/K & \cite{Lide2005} \\
 $\alpha_\text{ML}$ & 2.6e-6 $K^{-1}$ & \cite{Okada1984} \\
 $E_\text{ML}$ & 66.4 GPa & \cite{Knobloch2022} \\
 $\nu_\text{ML}$ & 0.19 & \cite{Knobloch2022} \\
 \hline
 $\rho_\text{Si}$ & 2.33 g/cc & \cite{Lide2005} \\
 $c_\text{p Si}$ & 712 J/kg/K & \cite{Lide2005} \\
 $\kappa_\text{Si}$ & 149 W/m/K & \cite{Lide2005} \\
 $\alpha_\text{Si}$ & 2.6e-6 $K^{-1}$ & \cite{Okada1984} \\
 $E_\text{Si}$ & 160 GPa & \cite{Hopcroft2010} \\
 $\nu_\text{Si}$ & 0.22 & \cite{Hopcroft2010} \\
 \hline
\end{tabular}
\caption{\textbf{Finite element model material parameters.} Values for nickel (Ni), the metalattice (ML) and silicon (Si), including porosity ($\phi$), density ($\rho$), specific heat capacity ($\text{c}_\text{p}$), thermal conductivity ($\kappa$), coefficient of thermal expansion ($\alpha$), Young's modulus (E) and Poisson's ratio ($\nu$).}
\label{tab1}
\end{table}

The FEA model also includes an 11-19 nm silicon overlayer between the nickel and metalattice, as observed in cross-sectional SEM measurements \cite{Knobloch2022}. This layer results from the high-pressure chemical vapor deposition process used to re-infiltrate silicon and reduce the metalattice porosity after etching of the silica template. It is not distinct from the metalattice itself, because it is annealed along with the metalattice, and is therefore crystalline and conforming to the FCC structure underneath. The thermal properties of this layer are uncertain due to its thinness and the lack of a clear boundary between it and the metalattice. We set its density, specific heat capacity, coefficient of thermal expansion and elastic properties to bulk crystalline silicon values. Its thermal conductivity is fit to the same value as the metalattice. While this may distort the fitted metalattice thermal conductivity at early times, we do not expect such a thin layer to have a major impact over the 8 ns fit window.

Five nickel grating geometries of varying linewidth, $\text{L}_\text{Ni}$, and periodicity, $\text{P}_\text{Ni}$, are measured experimentally and considered in the FEA model, as shown in Table \ref{tab2}. The five gratings will be hereafter referred to as 50-200, 100-400, 100-800, 500-2000 and 1000-4000, such that 50-200 describes the grating with $\text{L}_\text{Ni}$ = 50 nm and $\text{P}_\text{Ni}$ = 200 nm and so on. The $\text{L}_\text{Ni} / \text{P}_\text{Ni}$ ratio is held constant at 25\%, except for one 12.5\% case meant to test whether this ratio affects the conductivity fit. While the nickel grating period is set very precisely during the electron beam lithography process, the linewidth and height may vary, and are characterized with AFM. Cross-sectional SEM provides the silicon overlayer and metalattice film thicknesses, which are found to vary slightly between nickel gratings. These thickness variations are a known artifact of the metalattice fabrication process \cite{Knobloch2022}, do not imply a change in the metalattice physical properties, are accounted for in the FEA model and do not significantly affect the fitted thermal conductivity results. Table \ref{tab2} summarizes the various dimensions used as inputs in the FEA model for each nickel grating geometry.

\begin{table}[hbt!]
\centering
\begin{tabular}{ |p{1.5cm}|p{1.3cm}|p{0.8cm}|p{1.4cm}|p{1.5cm}|p{1.5cm}| }
 \hline
 $\textbf{Grating}$ & $\textbf{L}_\text{Ni}$ \newline [nm] & $\textbf{P}_\text{Ni}$ \newline [nm] & $\textbf{H}_\text{Ni}$ \newline [nm] & $\textbf{ML}$ \newline $\textbf{thickness}$ [nm] & \multirow{2}{1.5cm}{$\textbf{Si}$ \newline $\textbf{overlayer}$ [nm] } \\
 \hline
 50-200 & 51$\pm$6 & 200 & 11.7$\pm$0.3 & 583$\pm$31 & 11$\pm$5 \\
 100-400 & 106$\pm$8 & 400 & 11.7$\pm$0.3 & 555$\pm$2 & 12$\pm$4 \\
 100-800 & 106$\pm$7 & 800 & 11.7$\pm$0.3 & 503$\pm$54 & 14$\pm$7 \\
 500-2000 & 487$\pm$10 & 2000 & 11.7$\pm$0.3 & 450$\pm$56 & 17$\pm$9 \\
 1000-4000 & 989$\pm$15 & 4000 & 11.7$\pm$0.3 & 398$\pm$4 & 19$\pm$2 \\
 \hline
\end{tabular}
\caption{\textbf{Nickel grating geometries}. Nickel linewidth ($\text{L}_\text{Ni}$), period ($\text{P}_\text{Ni}$), height ($\text{H}_\text{Ni}$), metalattice thickness and silicon overlayer thickness for each of the five nickel grating geometries considered in the experiment and FEA model. All values are in nanometers. The nickel linewidth and height are measured with AFM and the error is a 90\% confidence interval constructed based on measurements at several locations on the sample. The nickel grating period is set very precisely by the electron beam lithography process. The metalattice and silicon overlayer thicknesses come from cross-sectional SEM measurements of the 100-400 and 1000-4000 gratings, and the error is the standard deviation of measurements at several points on the SEM image. The thickness values for the other gratings are interpolations based on where the grating is located on the sample surface relative to the imaged gratings and include error associated with this interpolation.}
\label{tab2}
\end{table}

Using the material parameters and geometries described above, the FEA model calculates temperature using Fourier's law combined with an energy balance equation,
\begin{eqnarray}\label{heatequation}
    &&\vec{q} = -\kappa \nabla T, \nonumber\\
    &&\rho \text{c}_\text{p} \frac{\partial T}{\partial t} + \nabla \cdot \vec{q} = \mathcal{Q}, 
\end{eqnarray}
where $T$ is temperature, $\vec{q}$ is heat flux, $\kappa$ is thermal conductivity, $\rho$ is density, $\text{c}_\text{p}$ is specific heat capacity, and $\mathcal{Q}$ is a heat source term. The calculation occurs over an 8 ns window after the injection of energy into the nickel structure via a simulated laser pulse. Since the nickel structures are thin and thermalize quickly, the laser pulse can be represented as a spatially uniform injection of heat into the nickel structure of the form
\begin{equation}\label{FEAPulse}
    \mathcal{Q}(t) = \mathcal{Q}_0 e^{-t/\tau_\text{ep}},
\end{equation}
where $\tau_\text{ep}$ = 454 fs is the characteristic electron-phonon equilibration timescale in nickel, as measured experimentally after an ultrafast laser excitation in a film of similar thickness to the gratings studied here \cite{Longa2007}. Because the penetration depth of the 800 nm pump light is approximately 13 nm in nickel and 8 $\upmu$m in silicon, we can safely assume $\mathcal{Q} = 0$ in the metalattice film and silicon substrate.

The nickel-silicon interface has a thermal boundary resistance (TBR) which must be accounted for in the FEA model. Previous works have experimentally measured nickel-silicon TBRs of 5 nK$\text{m}^2$/W \cite{Hopkins2015} and 2 nK$\text{m}^2$/W \cite{Frazer2019}, \cite{BeardoKnobloch2021}. While the TBR significantly affects the modeled thermal decay when the substrate is crystalline silicon \cite{Frazer2019}, the much smaller metalattice thermal conductivity dominates the TBR as the main bottleneck to thermal relaxation in the current study, making knowledge of the exact TBR less important. To account for variability in the fabrication and the rougher metalattice surface, we allow the TBR to vary between 2 and 8 nK$\text{m}^2$/W in the FEA model when calculating error bars on the metalattice thermal conductivity fits.

From the temperature profile, elastic properties, and coefficient of thermal expansion, the FEA model calculates a displacement vector $\vec{u}$ as a function of position across the sample surface. This calculation is done in the less computationally expensive quasi-static approximation, meaning that $\partial^2 \vec{u} / \partial t^2 = 0$, such that no acoustic waves are launched and the displacement decays monotonically after excitation. We use a home-built diffraction code to calculate the change in diffracted EUV intensity $\Delta I$ as a result of the displaced sample surface in the Fresnel approximation,
\begin{eqnarray}\label{Diffraction}
    &&\Delta I(x,t) = I(x,t) - I_0(x) = \nonumber\\
    &&\qquad |\mathcal{F}\{U(\xi,t)e^{ik\xi^2/2z}\}|^2 - |\mathcal{F}\{U_0(\xi)e^{ik\xi^2/2z}\}|^2,
\end{eqnarray}
where $\mathcal{F}$ denotes a Fourier transform, $\xi$ is the coordinate on the sample surface, $x$ is the coordinate on the camera chip, $z$ is the distance to the camera and $U(\xi,t)$ is the scalar electric field profile on the displaced sample surface. The profile for a static nickel grating $U_0(\xi)$ is approximated as
\begin{eqnarray}\label{grating}
    &&U_0(\xi) = \nonumber\\
    &&r_S + (r_G e^{-i \theta} - r_S) \text{comb}_{\text{P}_\text{Ni}}(\xi) \star \text{rect}(\frac{\xi}{\text{L}_\text{Ni}}),
\end{eqnarray}
where $r_S$ and $r_G$ are the complex substrate and grating reflectivities, $\text{L}_\text{Ni}$ and $\text{P}_\text{Ni}$ are the nickel grating linewidth and period, and $\theta$ is a phase accounting for the change in optical path length due to the grating height. The complex reflectivities $r_S$ and $r_G$ are calculated using an in-house EUV reflectivity solver based on the Parratt formalism \cite{Parratt1954,Esashi2021}, together with atomic scattering factors tabulated by Center for X-ray Optics (CXRO) \cite{Henke1993} to predict refractive indices of the materials at EUV wavelengths. All interfaces were assumed to be ideal with no roughness. The simulated change in EUV diffraction efficiency as a function of time can be summed over $x$ and directly compared to the experimental signal.

\section{Thermal Conductivity Fits and Error Bars}

The experimental EUV scatterometry data consists of 47 scans, which measure the EUV diffraction efficiency in Eq. \eqref{Diffraction} as a function of infrared pump-EUV probe delay time. The scans sample all five nickel grating geometries and also several different delay time windows and step sizes. Table \ref{tab3} summarizes the experimental EUV data used for the fits.

\begin{table}[hbt!]
\centering
\begin{tabular}{ |p{1.5cm}|p{2cm}|p{2cm}|p{1.5cm}| }
 \hline
 \textbf{Grating} & \textbf{Window (ps)} & \textbf{Step (ps)} & \textbf{Scans} \\
 \hline
 50-200 & 0-8000 & 40 & 6 \\
 50-200 & 0-750 & 5 & 4 \\
 50-200 & 0-1500 & 10 & 1 \\
 \hline
 100-400 & 0-8000 & 40 & 6 \\
 100-400 & 0-1500 & 10 & 6 \\
 \hline
 100-800 & 0-8000 & 40 & 4 \\
 100-800 & 0-3000 & 15 & 1 \\
 100-800 & 0-3500 & 20 & 2 \\
 100-800 & 0-4000 & 20 & 1 \\
 \hline
 500-2000 & 0-8000 & 40 & 8 \\
 \hline
 1000-4000 & 0-8000 & 40 & 8 \\
 \hline
\end{tabular}
\caption{\textbf{Experimental EUV data summary}. The apparent metalattice thermal conductivity is fit using 47 experimental scans, broken out here by nickel grating geometry, time window and temporal step size.}
\label{tab3}
\end{table}

Using the FEA model, we fit an apparent metalattice thermal conductivity to each of the above experimental EUV scans. The fits are done in MATLAB using the global search function, which uses an interior point method to find the apparent thermal conductivity value that minimizes the least squares error between simulated and experimental data. 

\subsection{Matrix Pencil Method}

Because the FEA simulations make the quasi-static approximation discussed above, they do not capture the surface acoustic waves present in the experimental data. In previous work on bulk silicon, it was possible to account for the surface acoustic waves via a computationally expensive FEA simulation of one or two oscillation periods without the quasi-static approximation \cite{BeardoKnobloch2021}. This procedure is impractical for the metalattice sample because its increased surface roughness relative to bulk silicon causes FEA simulations to overpredict the acoustic amplitude by a factor that varies between experimental scans. Instead we use the matrix pencil method (MPM) to remove surface acoustic waves from the experimental data and enable a direct comparison to the FEA simulations. The MPM algorithm begins by constructing a time-lagged covariance matrix from the experimental diffraction efficiency signal. It then performs a process known as singular value decomposition, wherein the covariance matrix is decomposed onto a basis that allows for separation of the time-varying components of the experimental signal from the time-invariant background noise. By fitting a complex exponential $e^{i \omega t}$ to each time-varying component and subtracting the oscillatory components, where Re($\omega$) $\neq$ 0, the MPM filters acoustics from the experimental signal, leaving only the monotonic quasi-static decay. The MPM algorithm can outperform more traditional methods like a fast Fourier transform in situations where oscillations are highly damped or partially obscured by noise. For more detail on the MPM algorithm see Supplementary Information section 3 in Ref. \cite{BeardoKnobloch2021}.

\subsection{Amplitude Considerations}

In addition to the thermal conductivity, it is necessary to fit an amplitude ratio between experiment and simulation, since normalizing both to the same value would overestimate the simulated amplitude by equating it with an experimental peak that is statistically likely to be inflated by noise. We therefore normalize the experimental scan to one, and fit the simulated amplitude. We observe that attempting to bound the amplitude fit within the approximate experimental noise is impractical; bounding within 1-2 standard deviations of the experimental noise eliminates too many legitimate fit outcomes, while bounding within $\geq3$ standard deviations is so lenient as to hardly be a constraint. Because of complications arising from the early-time inertial expansion, such as short-timescale acoustic modes within the nickel gratings themselves and uncertainty in the experimental time zero to within an experimental time step, we don't start fitting until the time at which experimental signal is maximized, which generally occurs before 200 ps.

\subsection{Error Bar Calculation}

To determine the uncertainty in the fitted thermal conductivity, we first confirm that certain potential sources of error have only a minor effect, including the EUV diffraction simulation parameters and the exact mesh parameters used in the FEA model. The sources of error which do matter are classified into three types: \textit{\textbf{model error}}, or variation in the fitted conductivity due to certain FEA model parameters, \textit{\textbf{experimental error}}, or variation between experimental data scans and \textit{\textbf{fit error}}, or variation due to the exact time window over which the conductivity is fit.

\begin{figure}
    \centering
    \includegraphics[width=0.9 \columnwidth]{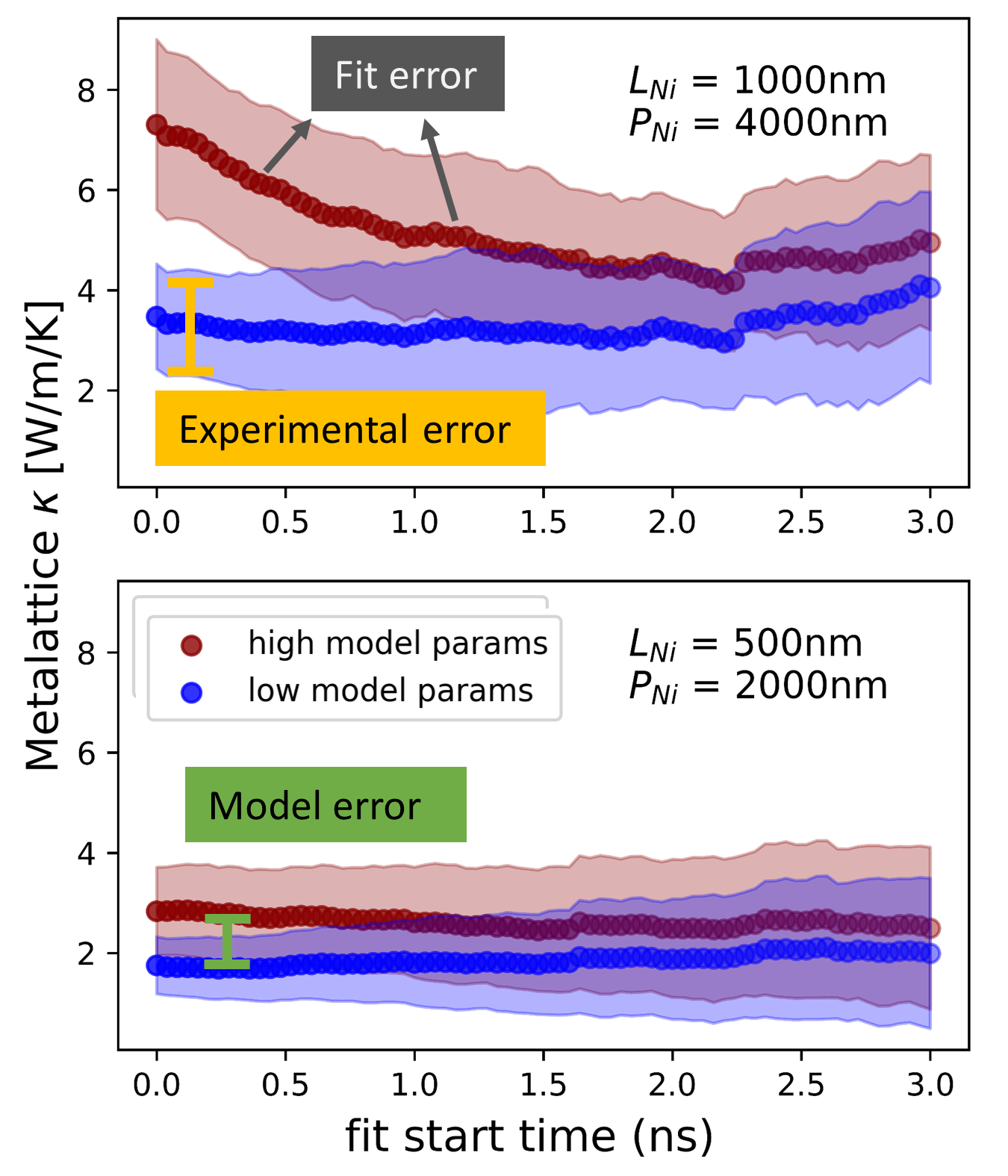}
    \caption{\textbf{Model error, experimental error and fit error.} Fitted apparent thermal conductivity vs. fit start time for the 1000-4000 (top) and 500-2000 (bottom) grating geometries. While the 500-2000 fit values are relatively independent of fit start time, the 1000-4000 values trend downward. The red and blue curves show the fitted conductivity for FEA simulations with parameters $\text{L}_\text{Ni}$, $\text{H}_\text{Ni}$, $\phi$, $\alpha_\text{ML}$ and TBR set within their bounds to produce maximal and minimal fit values respectively. The shaded regions show the experimental error due to variation between the 8 experimental data scans collected for each grating geometry.}
    \label{fig:FigureS2}
\end{figure}

The model error accounts for uncertainty in five FEA model parameters: $\text{L}_\text{Ni}$, $\text{H}_\text{Ni}$, $\phi$, $\alpha_\text{ML}$ and the thermal boundary resistivity (TBR) between the nickel grating and underlying silicon. The bounds on $\text{L}_\text{Ni}$, $\text{H}_\text{Ni}$ and $\phi$ come from the AFM measurements summarized in Table \ref{tab2} and $\phi=0.385 \pm 0.02$ from Ref. \cite{Knobloch2022}. Due to experimental evidence that the coefficient of thermal expansion for porous silicon can differ from that of bulk \cite{Faivre2000}, especially at high porosity, we bound $\alpha_\text{ML}$ = 2.6e-6 K$^{-1}$ $\pm$ 10\%. The TBR is the largest source of modeling error; we bound it between 2 and 8 nK$\text{m}^2$/W \cite{Frazer2019,Hopkins2015}, as discussed in Section \ref{MLFEA}. Other FEA model parameters do not significantly affect the fitted metalattice conductivity within physically reasonable bounds. For each nickel grating geometry, we determine the maximum and minimum conductivity fit achievable by varying the five relevant parameters within their bounds and set the model error to half the resulting discrepancy. The model error is illustrated as the distance between the red and blue curves in Fig. \ref{fig:FigureS2}.

The experimental error $\sigma_\text{E}$ can be calculated as
\begin{equation}\label{sigE}
    \sigma_\text{E} = \frac{S_n t_{(0.05,n-1)}}{\sqrt{n}}
\end{equation}
where $S_n$ is the sample standard deviation in fitted conductivity over $n$ experimental scans, and $t_{(0.05,n-1)}$ is the tabulated t-value for a two-sided 90\% confidence interval with $n-1$ degrees of freedom. The experimental error is illustrated by the red and blue shaded regions in Fig. \ref{fig:FigureS2}.

Fit error is the standard deviation of the variation in fitted conductivity with the temporal window over which the fit occurs. For most gratings, variations in conductivity with fit window are seemingly random and when aggregated yield a relatively consistent fitted conductivity regardless of whether the fit starts near the experimental peak or after several nanoseconds. As illustrated in the bottom panel of Fig. \ref{fig:FigureS2} for the 500-2000 grating, a time-independent apparent thermal conductivity fit confirms that heat flow in the metalattice obeys a Fourier law, since non-Fourier behavior would manifest itself in a time-dependent apparent thermal conductivity \cite{BeardoKnobloch2021}. However, for the 1000-4000 grating, the average fitted conductivity decreases as fit start time increases, as shown in the top panel of Fig. \ref{fig:FigureS2}. We observe a similar trend for the four long (8 ns) experimental scans on the 100-800 grating. These trends suggest a decay timescale that is not captured by the FEA model. However, we do not observe similar behavior for the other 35 out of 47 experimental scans. We therefore hypothesize that the trends are either caused by the higher experimental noise in measurements on the 1000-4000 and 100-800 gratings or by transient effects related to the silicon overlayer discussed in Section \ref{MLFEA}. We omit the experimental scans where fitted conductivity depends on fit start time when calculating the apparent metalattice thermal conductivity value presented in the main text.

The total uncertainty in apparent metalattice thermal conductivity $\sigma_\text{tot}$ is
\begin{equation}\label{sigE}
    \sigma_\text{tot} = \sqrt{\sigma_\text{E}^2 + \sigma_\text{M}^2 + \sigma_\text{F}^2},
\end{equation}
where $\sigma_\text{E}$, $\sigma_\text{M}$ and $\sigma_\text{F}$ are the experimental error, model error and fit error discussed above. Table \ref{tab4} summarizes the fitted thermal conductivity and error bars by grating and across the entire dataset. All fitted conductivity values agree within error bars except for 1000-4000, which does not agree with the smaller gratings. We believe that the higher fit for 1000-4000 is untrustworthy due to the variation in fitted conductivity with fit start time observed for this grating. Finally, when averaging the fitted conductivity over multiple experimental scans, we weight each scan by its signal-to-noise ratio, calculated as the ratio between the exponentially decaying and temporally uncorrelated components from the MPM algorithm.

\begin{table}[hbt!]
\centering
\begin{tabular}{ |p{2cm}|p{5.5cm}| }
 \hline
 \textbf{Grating} & \textbf{Fitted Conductivity W/m/K} \\
 \hline
 50-200 & 1.0 $\pm$ 0.3  \\
 100-400 & 1.5 $\pm$ 0.5 \\
 100-800 & 1.5 $\pm$ 0.6 \\
 500-2000 & 2.2 $\pm$ 1.1 \\
 1000-4000 & 4.2 $\pm$ 1.9 \\
 \hline
 \textbf{Aggregate} & \textbf{1.6 $\pm$ 0.4}  \\
 \hline
\end{tabular}
\caption{\textbf{Fitted thermal conductivity of the silicon metalattice.} Each experimental scan is fit separately, and results are aggregated by nickel grating and over the entire dataset. The final row excludes the 12 experimental scans for which fitted conductivity declines with fit start time.}
\label{tab4}
\end{table}

\begin{figure}
    \centering
    \includegraphics[width=0.9 \columnwidth]{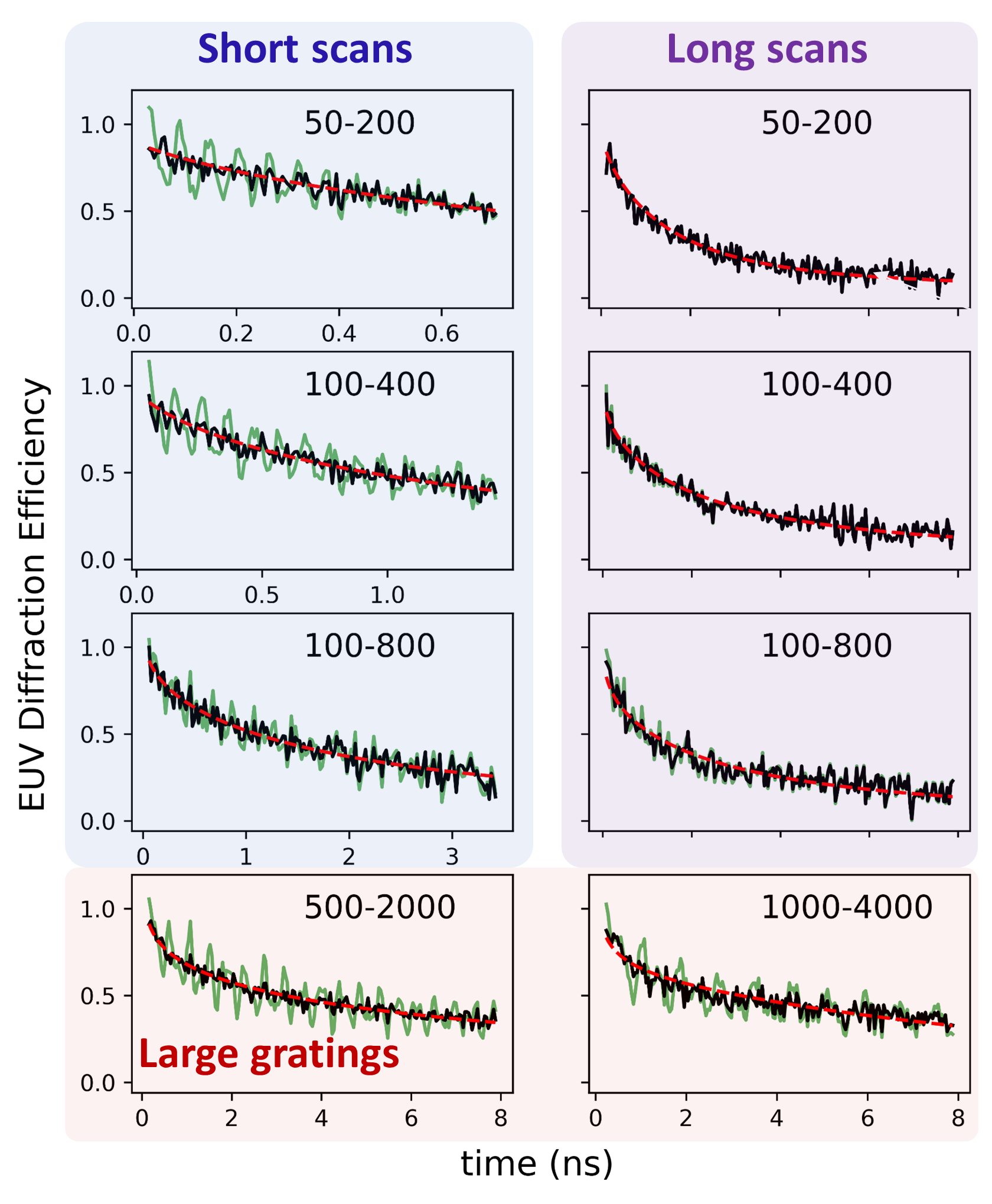}
    \caption{\textbf{Thermal conductivity fit quality.} Fits are aggregated by nickel grating geometry and scan length. The light green curve is the raw experimental data, the black curve is the MPM-filtered experimental data and the red curve is the best fit FEA simulation with parameters chosen to maximize the fitted conductivity. The fit quality is similar when FEA simulation parameters are chosen to minimize the fitted conductivity. The shorter scans on the smaller nickel gratings (shaded blue on the left side) were designed to sample the surface acoustic wave oscillations, whose wavelength is set by the grating period, and were used to extract the metalattice porosity in Ref. \cite{Knobloch2022}.}
    \label{fig:FigureS3}
\end{figure}

Finally, Fig. \ref{fig:FigureS3} shows the experimental scans and best fits aggregated by nickel grating geometry and scan length. The light green data are the raw experimental scans, while the black data are the MPM-filtered scans without the surface acoustic mode. The fit quality is generally good, with the exception of the long 8 ns 100-800 scans, in agreement with the above observation that fitted conductivity varies with fit start time in these measurements.   

\section{Derivation of a Darcy's Law Analog from the Guyer-Krumhansl Equation}\label{DarcyDerivation}

In this section, we describe how to obtain the Darcy's law analog
\begin{equation} \label{Darcy}
    \nabla T=-\frac{\mu}{K}\vec{q},
\end{equation}
along with an explicit expression for the permeability $K$ from the Guyer-Krumhansl equation (GKE):
\begin{equation}\label{GKE}
\tau\frac{\partial \vec{q}}{\partial t}+\vec{q}=-\kappa_\text{GK}\nabla T+\ell^2(\nabla^2 \vec{q}+\alpha_\text{GK}\nabla\nabla\cdot \vec{q}),
\end{equation}
where $\tau$, $\kappa_\text{GK}$, $\alpha_\text{GK}$, and $\ell$ are the heat flux relaxation time, the thermal conductivity, the volume viscosity coefficient, and the non-local length, respectively. In the confined phonon transport regime considered here, these unknown properties are expected to be geometry-dependent. According to the GKE, the inhomogeneity in the heat flux profile due to the presence of the pores causes a local reduction of the heat flux that diminishes the energy flowing throughout the structure, \textit{i.e.}, an apparent conductivity reduction. 

We first assume that heat transport in the metalattice is dominated by the viscous effects arising from the boundaries, in analogy to the Stokes regime for incompressible fluids at low Reynolds numbers. Accordingly, the heat flux term, the heat flux time derivative term, and the volume viscosity term appearing in Eq. \eqref{GKE} can be neglected with respect to the Laplacian term, due to the dominant boundary effects in the metalattice \cite{Alvarez2010, Karniadakis2005}. This assumption is microscopically justified using the heat flux profiles obtained from MD simulations in Section \ref{MolecularDynamicsB} below. Therefore, we obtain a Stokes equation analog in the absence of volume viscosity effects:
\begin{equation}\label{Stokes}
    \nabla T=\frac{\ell^2}{\kappa_\text{GK}}\nabla^2\vec{q}\equiv\mu\nabla^2\vec{q}.
\end{equation}

Now, we use the previous transport equation and appropriate boundary conditions to derive Darcy's law and an analytic expression for the permeability. The procedure is analogous to the Kozeny-Carman derivation \cite{Carman1956} along with the Klinkenberg quasi-ballistic correction \cite{Klinkenberg1941} for fluid transport in porous media. The phonon gas flows through interconnected tortuous passages of varying cross sections inside the nanostructured material. These passages can be approximated as a sequence of tubes of different sizes and orientations. We start by calculating the mean heat flux in a straight tube of diameter $D$ and length $L$ in a steady-state situation, where $D$ is understood to measure only the crystalline region and not include any disordered or amorphous layer between channel and pore \cite{Minnich2014}. We assume a homogeneous temperature gradient along the tube $\nabla T=-\Delta T/L$ and axisymmetric heat flow. We next combine the transport Eq. \eqref{Stokes} and the steady-state energy conservation equation ($\nabla\cdot \vec{q}=0$) to obtain,
\begin{equation}
    \frac{1}{r}\frac{\partial}{\partial r}\bigg(r\frac{\partial q}{\partial r}\bigg)=-\frac{1}{\mu}\frac{\Delta T}{L},
\end{equation}
whose solutions read
\begin{equation}\label{StokesSolutions}
    q(r)=-\frac{\Delta T}{L}\frac{r^2}{4\mu}+c_1\ln(r)+c_2,
\end{equation}
where $r$ is the radial coordinate. The integration constant $c_1=0$ since $q(r=0)$ is finite. In the standard derivation of the Poiseuille equation for fluid flow in a pipe, $c_2$ is obtained by imposing no-slip boundary conditions in the walls $q(r=D/2)=0$. Such no-slip condition is not adequate to describe the rarefied phonon gas. Even when assuming that the phonons collide diffusely at the pore walls, there is a significant heat flux at the boundary flowing along the direction of the thermal gradient. This slip flux emerges due to the phonon mean free paths being comparable to  $D$. 

Here we assume diffusive phonon-boundary collisions as expected for rough pore walls with defects comparable to the phonon wavelengths. Hence, the velocity direction of the phonons emitted from the boundary is randomly distributed, and thus the averaged contribution to the total heat flux of the emitted phonons is null. Conversely, the boundary-incident phonons experienced their previous collision somewhere inside the system. Their average velocity will thus correspond to the non-equilibrium conditions of the phonon gas layer where the previous collision took place. The distance between this layer and the boundary (denoted here as the slip length $\ell_\text{slip}$) can be estimated as the average mean free path of the non-localized phonons. In case $\ell_\text{slip}<D$, a significant fraction of boundary-incident phonons is interpreted to have a non-null average velocity in the direction of the thermal gradient, which translates into a non-null flux on the boundaries. It is worth noting that purely ballistic phonons travelling between opposite pore walls without scattering, if there are any, do not contribute to the slip flux, since they were randomly injected into the system. 

To proceed, we assume that the heat flux changes linearly with position close to the boundary. We can then write the slip flux boundary condition by estimating the contribution of the incident phonons as
\begin{equation}\label{slip}
q(r=D/2)=-C\ell_\text{slip} \frac{\partial q(r=D/2)}{\partial r},
\end{equation}
where $C$ is a dimensionless specularity parameter. This boundary condition can be refined by considering higher order corrections, which may account for microscopic boundary features such as surface disorder \cite{Karniadakis2005}. The parameter $C$ can be used to model a certain fraction of specular collisions \cite{Ziman2001}, or a certain fraction of pure ballistic phonons. Here we assume $C$=1 corresponding to fully diffusive phonon-boundary scattering.
The slip effect predicted by the boundary condition in Eq. \eqref{slip} is in perfect analogy to previous work discussing ballistic fluid transport in porous media \cite{Klinkenberg1941}. Eq. \eqref{slip} is satisfied if:
\begin{equation}
    c_2=\frac{\Delta T}{L}\frac{D^2}{16\mu}-\ell_\text{slip}\frac{\partial q(r=D/2)}{\partial r}.
\end{equation}
The spatial derivative in the previous equation can be obtained by taking the derivative of Eq. \eqref{StokesSolutions} with respect to $r$:
\begin{equation}
   \frac{\partial q(r=D/2)}{\partial r}=-\frac{\Delta T}{L}\frac{D}{4\mu},
\end{equation}
and we finally have
\begin{equation}
    c_2=\frac{\Delta T}{4\mu L}\bigg(\frac{D^2}{4}+\ell_\text{slip} D\bigg).
\end{equation}

We thus obtain the local heat flux 
\begin{equation}
    q(r)=\frac{\Delta T}{4\mu L}\bigg(-r^2+\frac{D^2}{4}+\ell_\text{slip} D\bigg),
\end{equation}
and the relation 
\begin{equation}\label{PoiseuilleAnalogWithSlip}
    \frac{\Delta T}{L}
    =\frac{32\bar{q}\mu}{D^2+8 \ell_\text{slip} D},
\end{equation}
where $\bar{q}$ is the mean heat flux value in the pipe:
\begin{equation}
    \bar{q}=\frac{4}{\pi D^2}\int^{D/2}_0 2\pi r q(r)dr=\frac{\Delta T}{4\mu L}\bigg(\frac{D^2}{8}+\ell_\text{slip} D\bigg).
\end{equation}

Let us now imagine the nanostructured material as a set of straight tubes with diameter $D$ and length $L$ that are homogeneously distributed in space. The total surface area of the nanosystem cross-section is $A$, whereas the actual cross-section for the phonon gas flow (total cross-section of the tubes) is $A'$, and we have $A'/A=1-\phi$. The total energy flow is $Q=A q$, where $q$ is the heat flux averaged over the total nanosystem cross-section. Note that the heat flux $q$ is the variable appearing in the effective Fourier relation used for experimental modeling. Moreover, the total energy flow can also be expressed as $Q=A'\bar{q}$, and thus we have $\bar{q}=q/(1-\phi)$. Now one can consider that the silicon tubes are not straight, but are longer tortuous passages surrounding the empty pores. In this case, the mean heat flux in the tubes must be larger to obtain the same total energy flow. By defining $L'$ as the mean length of the tortuous passages, we have $\bar{q}=(\frac{L'}{L})^2 q/(1-\phi)$ \cite{Carman1956}. The resulting mean heat flux can be inserted into Eq. \eqref{PoiseuilleAnalogWithSlip} to obtain
\begin{equation}\label{PoiseuilleAnalogWithSlipKozenyCarman}
    \frac{\Delta T}{L}
    =\frac{32 {L'}^2\mu q}{L^2(1-\phi)(D^2+8 \ell_\text{slip} D)}.
\end{equation}

In the following subsections, we show how to identify the permeability from Eq. (\ref{PoiseuilleAnalogWithSlipKozenyCarman}) for each nanosystem studied in the main text. 

\subsection{3D Distributions of Pores: Metalattices, Nanowire Networks and Porous Nanowires}
\label{Permeability}

Consider a metalattice with pores of diameter $d$ and a total empty pore volume $V=n\pi d^3/6$ (where $n$ is the number of pores). The volume of the empty pores $V$ and the volume of solid passages $V'=n'\pi D^2 L'/4$ (where $n'$ is the number of passages) satisfies $V/V'=\phi/(1-\phi)$. Moreover, the surface area of the passages where the heat flows $S'=n'\pi D L'$ is approximately equal to the total surface area of the pores $S=n\pi d^2/\Psi$ ($S'\approx S$). The dimensionless correcting factor $\Psi$ is known as the sphericity \cite{McCabe1993}, and is defined as the ratio of the surface area of a sphere with the same volume as the given pore to the surface area of the actual pore. Including this correcting factor is necessary to successfully compare Molecular Dynamics (MD) calculations, which simulate perfect pore shapes, to experiments, which have irregular pores. The sphericity can also be used to account for the cubic shape of the pores in the MD simulations performed in the present work (see Section \ref{MolecularDynamics}). We note that the equivalence of the surface areas ($S'\approx S$) is not an exact assumption for extremely high porosities or in the nanowire network, where the pores are not fully surrounded by silicon. However, it is adequate to characterize the general geometrical relations between the silicon passages and the empty spaces.

Combining all the previous geometric considerations, we have the following relation between the diameter of the tubes and the diameter of the pores
\begin{equation}\label{Dd_relation3D}
    D=\frac{2(1-\phi) \Psi d}{3\phi}.
\end{equation}
By inserting Eq. \eqref{Dd_relation3D} into Eq. \eqref{PoiseuilleAnalogWithSlipKozenyCarman}, we obtain the Darcy's law (Eq. \eqref{Darcy}), with the following permeability
\begin{equation}\label{permeabilityFULL}
    K=\frac{(1-\phi)^2 \Psi d}{24\Gamma^2\phi}\bigg(\frac{(1-\phi) \Psi d}{3\phi}+4\ell_\text{slip}\bigg),
\end{equation}
where we denoted $\Gamma\equiv L'/L$ as the tortuosity parameter. 

Note that the permeability can also be expressed as a function of the interpore distance or neck size, which corresponds to the silicon tube diameter $D$, instead of the pore size $d$. To evaluate $K$ in the porous nanowires \cite{LinaYang2017}, we use the alternative expression in terms of $D$ because the interpore distance is well characterized. In the case of the nanowire networks, we evaluate $K$ using the porosity and pore diameter determined by the wire sizes and periodicity.

In general, the tortuosity parameter $\Gamma$ depends on the symmetry of the distribution of pores and the porosity, which determine the average length of the solid passages through the material. For simple-cubic pore distributions and 3D networks simulated using MD, we assume $\Gamma=1$ since the phonon gas can follow straight paths through the nanosystem. For experiments, we assume $\Gamma^2= 5/2$ as originally proposed by Carman to account for the tortuous paths the gas must trace in realistic samples exhibiting defects and pore geometry variability \cite{Carman1956}. We note that, in general, tortuosity might vary in the range $1<\Gamma^2<5/2$ depending on porosity and pore distribution \cite{Graczyk2020,Matyka2008}. The use of more accurate porosity-dependent tortuosity functions might be used to understand conductivity variations in some experiments in porous silicon \cite{Ferrando2018}. However, it does not significantly modify the estimated permeabilities in the different experiments considered here.

Finally, we assume that the slip length is geometrically determined by the tube diameter as $\ell_\text{slip}=D/4$. This assumption is justified in Section \ref{MolecularDynamicsB} by fitting the heat flux profile obtained in MD simulations of multiple geometries (see Fig. \ref{fig:FigS6-5}). The previous ansatz is used in Eq. \eqref{permeabilityFULL} to obtain the final expression for the 3D permeability (Eq. 5 in the main text). We note that more accurate estimations of the permeability might be obtained by carefully characterizing the slip length in each sample, which might account for specific surface disorder or backscattering effects. These refinements on  $\ell_\text{slip}$ characterization would directly modify the permeability according to Eq. \eqref{permeabilityFULL}.

\subsection{2D Distributions of Pores: Nanomeshes}

The reasoning discussed in the previous subsection can be used to derive the permeability in geometries with a 2D distribution of pores such as a nanomesh \cite{Heath2010}. In this case, the volume of the empty cylindrical pores is $V=nh\pi d^2/4$, where $d$ is the diameter of the pore and $h$ is the thickness of the membrane. The volume of solid passages is $V'=n'\pi D^2L'/4$ and the surface area is $S'=n'\pi DL'$, and the same relation $V/V'=\phi/(1-\phi)$ is satisfied. In the nanomeshes considered in this work, the membrane thickness is comparable to the interpore distance. Hence, the role of the top and bottom membrane boundaries cannot be neglected when computing the pore size, and we have $S=n(\pi d h+2(p^2-\pi d^2/4))/\Psi$, where $p$ is the periodicity of the nanomesh and the sphericity $\Psi$ is redefined as the ratio of the surface area of the nominal nanomesh unit cell to its actual surface area. As in the 3D case, by imposing $S=S'$, we obtain the geometric relation:
\begin{equation}
    D=\frac{(1-\phi)d^2h\Psi}{\phi(dh+2p^2/\pi-d^2/2)}.
\end{equation}
Now, using Eq. \eqref{PoiseuilleAnalogWithSlipKozenyCarman} and using again the slip length ansatz $\ell_\text{slip}=D/4$, we obtain the nanomesh permeability:
\begin{equation}\label{permeabilityNanomesh}
    K=\frac{3d^4h^2(1-\phi)^3\Psi^2}{32\Gamma^2 \phi^2(dh+2p^2/\pi-d^2/2)^2}.
\end{equation}
The previous expression reduces to the standard result for fluids \cite{Wagner2021} if the ballistic correction and the influence of the top and bottom boundaries are neglected. Additionally, we assume the same tortuosity factor $\Gamma^2=5/2$ used in the other experiments. However, slightly lower tortuosity values might be considered in this case because the pores are aligned. This would translate to proportionally higher values of the apparent viscosity. 

\subsection{Application to Specific Experimental and MD Geometries}\label{DarcyLawC}

\begin{figure}
    \centering
    \includegraphics[width=0.98 \columnwidth]{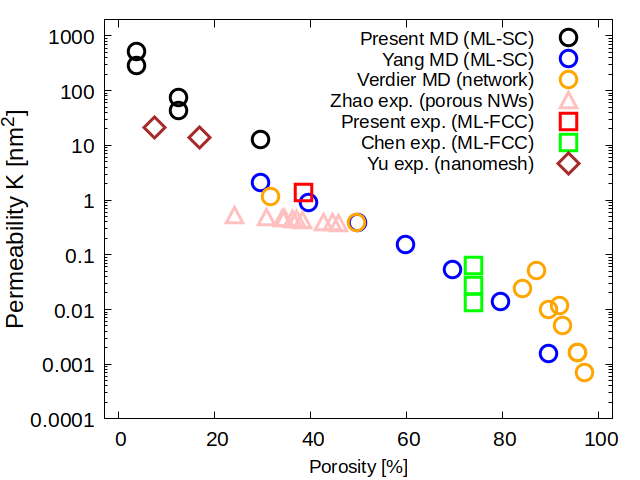}
    \caption{\textbf{Permeability $K$ as a function of porosity for the different nanosystems considered.}}
    \label{fig:FigureS5}
\end{figure}

In this subsection we detail the geometric features and resulting permeability of the systems considered in Fig. 2 of the main text. The MD-simulated metalattices (black circles) consist of a simple cubic (SC) distribution of cubic pores with periodicity of 13.03 nm and pore side ranging from 8.69 to 4.34 nm, and with periodicity 9.78 nm  and pore size ranging from 4.89 to 3.26 nm (see Section \ref{MolecularDynamics}). These results are compared to previous simulations of higher porosity metalattices (blue circles), consisting of a SC distribution of spherical pores with lattice constant 4.34 nm and varying pore size \cite{Baowen2014}, and to simulations of 3D networks (yellow circles), where the wire diameter and the periodicity are varied from 1.63 nm to 4.88 nm, and 4.35 nm to 27.12 nm, respectively \cite{Termentzidis2018}. The present experiment (red square) considers a metalattice with a face-centered-cubic (FCC) distribution of pores with 36 nm periodicity and 20 nm pore diameter. Previous measurements allow a comparison to similar metalattice films (green squares) with $\phi = 0.75$ and a close-packed FCC distribution of pores with radii of 7, 10, and 15 nm \cite{Dabo2020}. In both experimental metalattice samples, the sphericity is $\Psi=0.36$, as characterized from electron tomography images of the lower-porosity sample (see Section \ref{Fabrication}). In the nanomesh experiments (brown diamonds), the cylindrical pore diameter varies from 11 to 16 nm, with fixed membrane thickness of 22 nm and periodicity of 34 nm \cite{Heath2010}. The sphericity is $\Psi=0.75$, as characterized from scanning electron microscope images (see Fig. 1(b) in Ref. \cite{Heath2010}). Finally, in the porous nanowire experiment (pink triangles), the porosity is varied while maintaining an interpore distance $D=$ 4.3 nm \cite{LinaYang2017}. In this case neither $d$ nor $\Psi$ are required to evaluate the permeability (see Eq. \eqref{Dd_relation3D}). In the MD simulations, where the pores are perfectly aligned, we assume the minimum tortuosity $\Gamma=1$, while in the experiments, we assume $\Gamma^2=5/2$ to account for the tortuous paths in realistic samples with defects and geometric variability \cite{Carman1956}. In Fig. \ref{fig:FigureS5} we show the permeability values calculated for these systems as described in the previous subsections, which we use to compute the viscosity plotted in Fig. 2 of the main text.

We note some uncertainties that could slightly modify the permeability calculations and thus influence the comparison of the phonon viscosity $\mu$ in experiments and MD simulations. In the formulation of the permeability, we assumed purely diffusive phonon-boundary collisions, \textit{i.e.,} $C=1$ in Eq. \eqref{slip}. However, the slip length $\ell_\text{slip}$, characterized using atomistic simulations and discussed in Section \ref{MolecularDynamicsB} (see Fig. \ref{fig:FigS6-5}), might account for some degree of specularity in MD even at room temperature due to the atomically smooth boundaries. This is not expected in experiments, where the pore walls are rough, and would translate into a slightly smaller $\ell_\text{slip}$ in experiments, resulting in smaller values of the permeability (see Eq. \eqref{permeabilityFULL}), and proportionally smaller values of the viscosity. Another possible source of variation in $\mu$ is the hypothetical presence of amorphous layers surrounding the pores in experiments. These layers are not simulated in MD and might reduce the heat flux flowing through the structure \cite{DESMARCHELIER2022123003,Lysenko1999}, which would result in an increased viscosity. Moreover, the same sphericity characterized in the present metalattice sample (see Section \ref{Fabrication}) is assumed for previous experiments in similarly fabricated metalattices \cite{Dabo2020}. This might overlook some variation in $K$ between the two experiments, since it is reasonable to expect that the quality of the pores is distinct in each specific case, due to the additional silicon infiltration step necessary to produce the lower-porosity sample \cite{Knobloch2022}. It is also worth noting that a fixed pore spacing is assumed for all porosities in the porous nanowire system \cite{LinaYang2017}, but a slight increase in pore spacing is expected for the smallest porosity, which would result in an increased permeability. This would reduce the discrepancy between the estimated viscosity in this case and the scaling shown in Fig. 2 of the main text, consistent with other experimental measurements on porous nanowires of similar porosity \cite{Ferrando2018}, which display an average viscosity matching the universal trend of $\mu$. Finally, in the nanomesh experiment studied here \cite{Heath2010}, an upper bound on the apparent conductivity is reported, because the conductivity is estimated only accounting for the nonporous cross-sectional surface area. Using the exact apparent conductivity $\kappa$ in Eq. 4 of the main text would lead to an slight increase in the characterized viscosity in this case.

\section{Molecular Dynamics Simulations}\label{MolecularDynamics}

We use nonequilibrium MD simulations to obtain the apparent thermal conductivity and heat flux distributions in metalattices with a SC distribution of cubic pores. Three periodic supercells, each one including a pore, are simulated in steady-state. We use LAMMPS software \cite{PLIMPTON19951} with a Stillinger-Weber empirical interatomic potential \cite{SWpotential}. To avoid rotational or translational effects, a 1.09 nm-thick layer of fixed atoms is added to two opposite ends of the system. Using a Langevin thermostat and a time step of 0.5 fs, the rest of the system is first thermalized at 300 K over 1 ns. Then, a thermal gradient is imposed by thermalizing the layers of atoms in contact with the fixed ends using a Langevin thermostat at 310 K (source) and 290K (sink), respectively, while leaving the central atoms free. The thickness of the thermalized layers is 2.17 nm. Finally, we use periodic boundary conditions in the two directions orthogonal to the imposed thermal gradient. The steady-state is reached by running the simulation for 1 ns with a time step of 0.5 fs. Thereafter, the apparent thermal conductivity and heat flux profiles are averaged over the subsequent 4 ns using the same time step size. The results are averaged over two simulations in each case.

The free central atoms in between the heat sink and source consist of three periodic metalattice supercells (three pores). The cubic side length of a supercell consists of 24 or 18 conventional cells (8 atoms each), with a total length of 13.03 nm or 9.77 nm, respectively. The atoms within a cubic domain in the center of each supercell are removed to create the pore. The size of the pore is varied to simulate a variety of porosities. For the supercell with 24 conventional cell side length, pore side lengths of 8, 12 and 16 conventional cells (4.35, 6.52, 8.69 nm) are considered, to simulate metalattices with porosities of 0.037, 0.125 and 0.296, respectively. For the supercell with 18 conventional cell side length, pore side lengths of 6 and 9 conventional cells (3.26, 4.89 nm) are considered, with porosities of 0.037 and 0.125, respectively.

\begin{figure}
    \centering
    \includegraphics[width=0.9 \columnwidth]{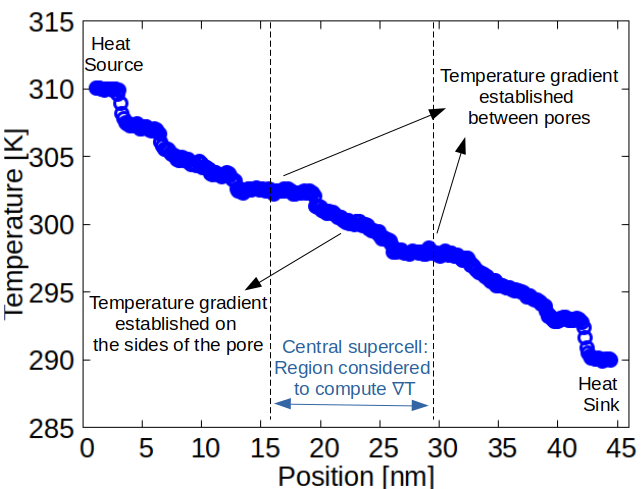}
    \caption{\textbf{Example temperature profile between thermostats.} This case corresponds to a metalattice with three supercells (three pores) between thermostats, with a supercell side length of 24 conventional cells and pore side length of 12 conventional cells ($\phi$=0.125).}
    \label{fig:FigS6-1}
\end{figure}

The average thermal gradient $\nabla T$ and the average heat flux $\vec{q}$ are required to calculate the apparent conductivity $\kappa=|\vec{q}|/|\nabla{T}|$. The temperature is defined using the kinetic energy of the atoms:
\begin{equation}
    T=\frac{m}{3 k_B N}\sum^N_i \vec{v}_i^2,
\end{equation}
where $m$ is the silicon atomic mass, $k_B$ is the Boltzmann constant, $\vec{v}_i$ is the velocity of atom $i$, and $N$ is the number of atoms averaged. To extract the temperature profile established between the thermostats (see Fig. \ref{fig:FigS6-1}), the temperature is locally averaged within layers of monoatomic thickness coplanar to the source and sink. To extract the average thermal gradient from the temperature profile, only the central supercell is considered, so as not to include the nonlinear effects in the temperature profile close to the thermostated layers. Moreover, the average heat flux is computed considering the kinetic and the virial contributions \cite{Hardy1963},
\begin{equation}\label{microscopicFlux}
    \vec{q}=\frac{1}{V_\text{T}}\sum^N_i \big(e_i\vec{v}_i+\bar{\bar\sigma}_i\cdot \vec{v}_i\big),
\end{equation}
where $e_i$ is the total energy of atom $i$, $\frac{N}{V_\text{T}}\bar{\bar\sigma}_i$ is the stress tensor on atom $i$, and $N$ is the number of atoms within the volume $V_\text{T}$. This volume only includes the central supercell. However, since we are characterizing the apparent conductivity of the metalattice, the integration volume $V_\text{T}$ includes both the silicon and the empty pore regions.

\begin{figure}
    \centering
    \includegraphics[width=1. \columnwidth]{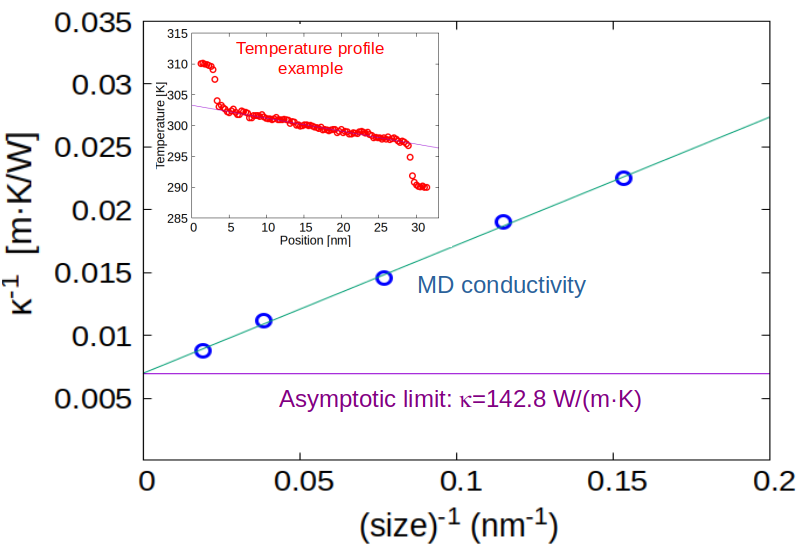}
    \caption{\textbf{Asymptotic calculation of the bulk thermal conductivity.} The inverse of the thermal conductivity $\kappa^{-1}$ is plotted as a function of the inverse of the system size (distance between thermostats) for a system with no pores. The asymptotic value of the conductivity is 148.2 W/mK. The inset shows an example of the linear temperature profile between thermostats obtained in the simulations.}
    \label{fig:FigS6-2}
\end{figure}

The average heat flux can be alternatively obtained by measuring the energy introduced into the thermostats to maintain the temperature difference in steady state. The rate of energy injection (removal) $Q$ at the heat source (sink) corresponds to the energy that is flowing between thermostats. As expected, the rate $Q$ reaches a constant value in steady-state. Therefore, one can compute the average heat flux as:
\begin{equation}
    |\vec{q}|=Q/A_\text{T},
\end{equation}
where $A_\text{T}$ is the total cross-sectional area of the metalattice, accounting both for the silicon and porous regions. The discrepancy between the two methods of calculating the heat flux is consistently less than 5$\%$ for all the cases under consideration.  Moreover, we note that the average component of the heat flux parallel to the thermal gradient is almost two orders of magnitude larger than the average transverse components, whose values fluctuate around zero due to the periodic boundary conditions. 

To further validate the present MD simulations and the method used to extract the apparent metalattice thermal conductivity, we compute the bulk thermal conductivity of silicon. To do this, we use the same setup described above, without removing the central atoms (no empty pores). In this case, only the linear temperature profile is used to characterize the thermal gradient. As expected, long-range interactions between the thermostats produce a strong size effect \cite{Sellan2010}. Therefore, the bulk conductivity is extracted by asymptotic methods \cite{Schelling2002} (see Fig. \ref{fig:FigS6-2}). We obtain a bulk conductivity of 148.2 W/mK, very close to the real value at room temperature. 

Eq. \eqref{microscopicFlux} is also used to obtain the local heat flux distributions $\vec{q}$ in Fig. 3 of the main text and in Section \ref{MolecularDynamicsB} below. In this case, the integration volume $V_\text{T}$ is restricted to small cubic boxes of 2x2x2 conventional cell size (side length of 1.086 nm). The error bars of the local heat flux are estimated during the time averaging, resulting in a relative error of 7$\%$. For the 2D heat flux distribution shown in Fig. 3 of the main text, we use a linear interpolation of the MD data.

\subsection{Size Effects}

\begin{figure*}
    \centering
    \includegraphics[width=2. \columnwidth]{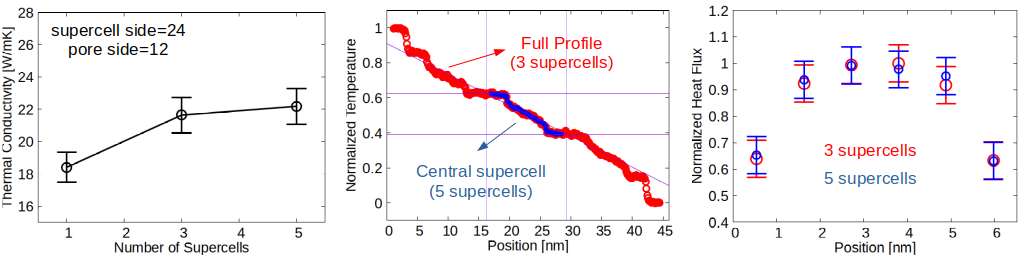}
    \caption{\textbf{Size effects study}. No significant change in thermal conductivity, thermal gradient or heat flux profile is obtained when increasing the number of supercells between source and sink from three to five. The supercell side length is 24 conventional cells and the pore side length is 12 conventional cells ($\phi=0.125$). The apparent thermal conductivity (left) becomes independent of system size above three supercells. Moreover, the normalized thermal gradient (center) and heat flux profile (right) calculated in the central supercell are not modified by increasing the system size from three to five supercells.}
    \label{fig:FigS6-3}
\end{figure*}

As shown in Fig. \ref{fig:FigS6-2}, the bulk conductivity estimation in MD is influenced by the distance between thermostats, and can only be extracted asymptotically. We verify that this size effect does not influence the conductivity and flux estimations in the metalattice simulations, which consider three supercells. To do this, we compare results simulating one, three and five supercells between source and sink (see Fig. \ref{fig:FigS6-3}). We first verify that the temperature and heat flux profiles in the central supercell are equivalent if considering three vs. five supercells between thermostats. Moreover, the observed change in thermal conductivity when increasing the number of supercells from 3 to 5 is within numerical error (smaller than 5$\%$). Therefore, the long-range interactions between thermostats leading to size effects in the bulk simulations are attenuated due to the presence of the pores, and simulating three supercells is sufficient to obtain converged results for the apparent metalattice conductivity and heat flux profiles. We also verified that neither the apparent conductivity nor the local heat flux profiles are significantly modified by increasing or slightly reducing the thickness of the heat source or heat sink.

\subsection{Guyer-Krumhansl and Stokes Fits to MD and Model Applicability}\label{MolecularDynamicsB}

\begin{figure*}
    \centering
    \includegraphics[width=2. \columnwidth]{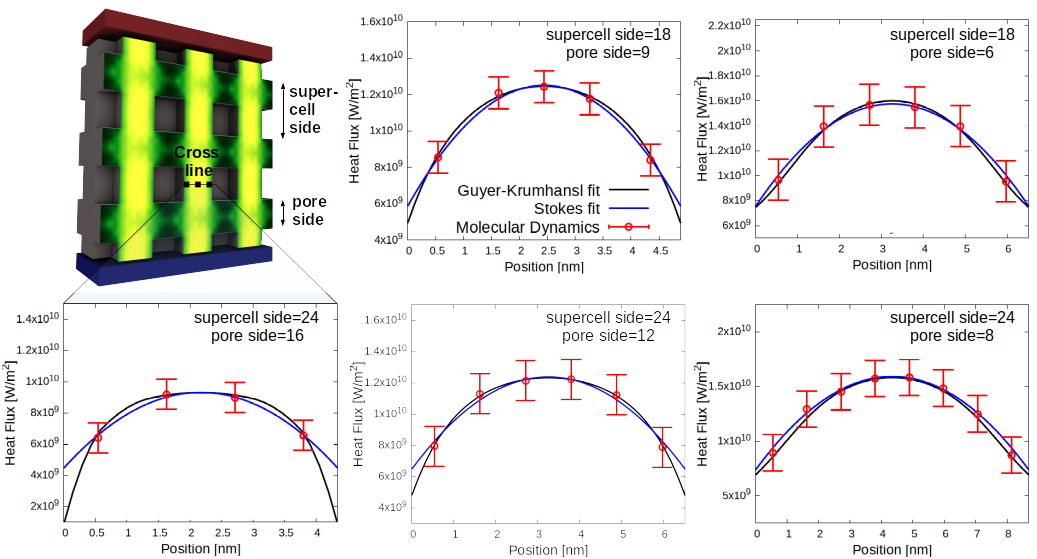}
    \caption{\textbf{Fluid-like heat flow in the metalattice}. The heat flux profile at the cross-line between two contiguous pores, as obtained in MD simulations (red) is plotted along with the corresponding Stokes (blue) and GKE (black) fits. Multiple cases are considered with various supercell and pore sizes. For each case, the supercell and pore side lengths are specified in terms of the number of conventional cells, which consist of 8 atoms in a cube of side length 0.5431 nm.}
    \label{fig:FigS6-4}
\end{figure*}

\begin{figure}
    \includegraphics[width=1. \columnwidth]{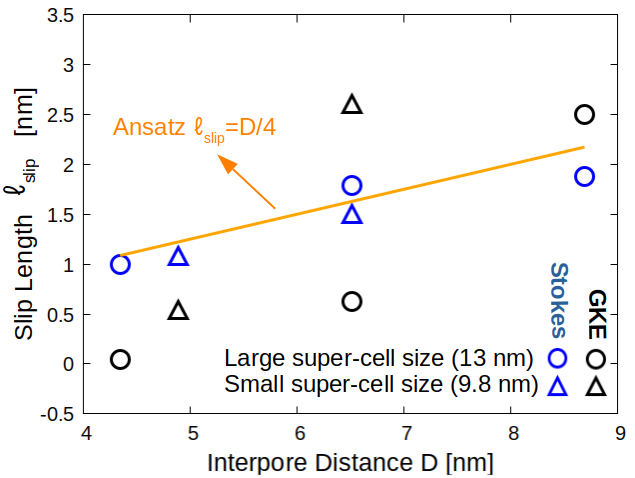}
    \caption{\textbf{Slip Length analysis}. Slip length values fitted to MD heat flux profiles using the GKE (black) and Stokes equation (blue). For the GKE, $\ell_\text{slip}$ is mainly determined by the porosity. For Stokes, $\ell_\text{slip}\sim D/4$, where $D$ is the interpore distance, as assumed when deriving the permeability expressions.}
    \label{fig:FigS6-5}
\end{figure}

In this subsection we describe the hydrodynamic modeling of the local heat flux profiles obtained from MD simulations. We fit the GKE in Eq. \eqref{GKE} and the Stokes approximation in Eq. \eqref{Stokes} to MD and validate the assumption of the slip length $\ell_\text{slip}=D/4$ used in the derivation of Darcy's law and the permeability.

As shown in Fig. 3 of the main text, the GKE with appropriate boundary conditions, namely thermal insulation ($\vec{q}\cdot \vec{n}=0$) and slip flow (Eq. \eqref{slip} with $C=1$), can be used to reproduce the heat flux distribution inside the metalattice. For completeness, in Fig. \ref{fig:FigS6-4} we show the GKE fits to MD along a cross-section for different sizes of the supercell and the pore. We also show the Stokes fits with the same boundary conditions, to locally reproduce the heat flux in the passages between the pores. The GKE fits are performed using COMSOL Multiphysics \cite{comsol,Beardo2019} in the exact geometry of the metalattice, including the pores. Conversely, the Stokes equation is used to model the heat flux in a tube with diameter equivalent to the interpore distance $D$. Even though the Stokes approximation is not appropriate to describe the flux profiles in the complex geometry, particularly near corners, it can locally reproduce the flux distributions in the interpore channels. This justifies the validity of the Stokes regime assumption in describing the energy flow at these scales, as required when deriving the Darcy's law analog (see Section \ref{DarcyDerivation}). 

The quality of the Stokes fit is expected to decrease as the interpore distance increases, due to the increasing influence of the heat flux term in the GKE. These deviations are expected to restrict the applicability of the Stokes equation (and consequently Darcy's law) to systems with small enough characteristic sizes. This breakdown of the Stokes approximation is not observed in the considered cases at room temperature, thus microscopically justifying the applicability of the model for nanostructured silicon samples with feature sizes of a few tens of nanometers, including the cases considered here (see Section \ref{DarcyLawC}). However, we note that for systems where heat flow is significantly less confined along at least one dimension (\textit{e.g.}, \cite{PeidongYang2017,Hopkins2010}), the viscosity characterized using $K$ and $\kappa$ is systematically larger than predicted by the scaling shown in Fig. 2 of the main text. This indicates that the applicability of the present analytical model is restricted to highly confined nanosystems.

\begin{table}[hbt!]
\centering
\begin{tabular}{ |p{1.75cm}|p{1.2cm}|p{0.9cm}|p{0.9cm}|p{1.2cm}|p{0.9cm}|p{0.9cm}| }
 \hline
 \textbf{Geometry} (supercell $\ \&$ pore size) & GKE $\kappa_\text{GK}$ [W/mK] & GKE $\ \ \ \ \ell\ \ \ \ $ [nm] & GKE  $\ell_\text{slip}$ [nm] & \textcolor{blue}{Stokes $\kappa_\text{GK}$ [W/mK]} & \textcolor{blue}{Stokes $\ \ \ \ \ell\ \ \ \ $  [nm]} & \textcolor{blue}{ Stokes $\ell_\text{slip}$ [nm]} \\
 \hline
  18 $\&$ 6 &  45 & 2.0 & 2.6 & \textcolor{blue}{6.1} & \textcolor{blue}{2.0} & \textcolor{blue}{1.5}  \\
  \hline
  18 $\&$ 9 & 24 & 0.9 & 0.54 & \textcolor{blue}{2.75} & \textcolor{blue}{0.9} & \textcolor{blue}{1.08} \\
  \hline
 24 $\&$ 8  & 69 & 2.5 & 2.5 & \textcolor{blue}{10.25} & \textcolor{blue}{2.5} & \textcolor{blue}{1.88}\\
  \hline
 24 $\&$ 12 & 34 & 1.05 & 0.63 & \textcolor{blue}{2} & \textcolor{blue}{1.05} & \textcolor{blue}{1.79}\\
  \hline
 24 $\&$ 16 & 20 & 0.5 & 0.05 & \textcolor{blue}{0.95} & \textcolor{blue}{0.5} & \textcolor{blue}{1.0}\\
 \hline
\end{tabular}
\caption{\textbf{GKE and Stokes parameter values fitted to the heat flux profiles obtained in MD simulations.} The geometry is defined in terms of the number of conventional cells on the side of the cubic supercell and the cubic pore.}
\label{tab4}
\end{table}

The GKE and Stokes fit parameters ($\ell^2,\kappa_\text{GK},\ell_\text{slip}$)  are displayed in Table \ref{tab4}. In the GKE modeling, both $\ell$ and $\kappa_\text{GK}$ are fit independently and determined unequivocally. In contrast, the same profile can be obtained in the Stokes modeling by modifying $\ell^2$ and $\kappa$, while maintaining fixed the ratio $\mu$. Here, we arbitrarily fixed $\ell$ to be equal to the GKE fits and modified $\kappa_\text{GK}$ to reproduce the MD profiles with the Stokes equation However, the slip length $\ell_\text{slip}$ is independent of this choice, {\it i.e.}, a single value of the slip length fits the profile in the GKE or Stokes fits, respectively. 

All fit parameters are dependent on the supercell and pore sizes. However, the ratio $\mu=\ell^2/\kappa_\text{GK}$ seems to be mainly dependent on porosity. As shown in Fig. 3 of the main text, this is consistent with the experimental characterization of the viscosity $\mu$. Finally, the fitted slip length is mainly determined by porosity $\phi$ in the GKE fits, and is linearly dependent on the interpore distance $D$ in the Stokes fits (see Fig. \ref{fig:FigS6-5}). Specifically, the slip length according to the Stokes fits can be approximated as $\ell_\text{slip}\sim D/4$. This justifies the ansatz assumed for this parameter in the theoretical derivation of Darcy's law in Section \ref{DarcyDerivation}. 

\newpage

\bibliographystyle{ieeetr}
\bibliography{biblio}